\documentclass[twocolumn]{aastex61}
\submitjournal{ApJ}
\shorttitle{Imaging the $R_s$-scale Structure of M87 with the EHT using Sparse Modeling}
\shortauthors{K.~Akiyama et al.}
\newcommand{\haystack}{Massachusetts Institute of Technology, Haystack Observatory, Route 40, Westford, MA 01886, USA}
\newcommand{\naoj}{Mizusawa VLBI Observatory, National Astronomical Observatory of Japan, 2-21-1 Osawa, Mitaka, Tokyo 181-8588, Japan}
\newcommand{\utokyo}{Department of Astronomy, Graduate School of Science, The University of Tokyo, 7-3-1 Hongo, Bunkyo-ku, Tokyo 113-0033, Japan}
\newcommand{\ism}{Department of Statistical Science, School of Multidisciplinary Sciences, Graduate University for Advanced Studies,  10-3 Midori-cho, Tachikawa, Tokyo 190-8562, Japan}
\newcommand{\sokendaiism}{Graduate University for Advanced Studies, 10-3 Midori-cho, Tachikawa, Tokyo 190-8562, Japan}
\newcommand{\sokendainaoj}{Department of Astronomical Science, School of Physical Sciences, Graduate University for Advanced Studies, 2-21-1 Osawa, Mitaka, Tokyo 181-8588, Japan}
\newcommand{\cfa}{Harvard Smithsonian Center for Astrophysics, 60 Garden Street, Cambridge, MA 02138, USA}
\newcommand{\perimeter}{Perimeter Institute for Theoretical Physics, 31 Caroline Street, North Waterloo, Ontario N2L 2Y5, Canada}
\newcommand{\uwaterloo}{Department of Physics and Astronomy, University of Waterloo, 200 University Avenue West, Waterloo, Ontario N2l 3G1, Canada}
\newcommand{\mpe}{Max Planck Institute for Extraterrestrial Physics, Giessenbachstr. 1, 85748 Garching, Germany}
\newcommand{\radboud}{Department of Astrophysics/IMAPP, Radboud University Nijmegen, P.O. Box 9010, 6500 GL Nijmegen, The Netherlands}
\newcommand{\csail}{Massachusetts Institute of Technology, Computer Science and Artificial Intelligence Laboratory, 32 Vassar Street, Cambridge, MA 02139, USA}
\newcommand{\utokyos}{Department of Astronomy, School of Science, The University of Tokyo, 7-3-1 Hongo, Bunkyo-ku, Tokyo 113-0033, Japan}
\begin{document}
\title{Imaging the Schwarzschild-radius-scale Structure of M87 with the Event Horizon Telescope using Sparse Modeling}
%
%
\correspondingauthor{Kazunori Akiyama}
\email{kazu@haystack.mit.edu}
\author{Kazunori Akiyama}
\altaffiliation{JSPS Postdoctoral Fellow for Research Abroad}
\affil{\haystack}
\affil{\naoj}
\affil{\utokyo}
\author{Kazuki Kuramochi}
\affil{\utokyo}
\affil{\naoj}
\author{Shiro Ikeda}
\affil{\ism}
\affil{\sokendaiism}
\author{Vincent L. Fish}
\affil{\haystack}
\author{Fumie Tazaki}
\affil{\naoj}
\author{Mareki Honma}
\affil{\naoj}
\affil{\sokendainaoj}
\author{Sheperd S. Doeleman}
\affil{\haystack}
\affil{\cfa}
\author{Avery Broderick}
\affil{\perimeter}
\affil{\uwaterloo}
\author{Jason Dexter}
\affil{\mpe}
\author{Monika Mo{\'s}cibrodzka}
\affil{\radboud}
\author{Katherine L. Bouman}
\affil{\csail}
\author{Andrew Chael}
\affil{\cfa}
\author{Masamichi Zaizen}
\affil{\utokyos}
%
\begin{abstract}
We propose a new imaging technique for radio and optical/infrared interferometry. The proposed technique reconstructs the image from the visibility amplitude and closure phase, which are standard data products of short-millimeter very long baseline interferometers such as the Event Horizon Telescope (EHT) and optical/infrared interferometers, by utilizing two regularization functions: the $\ell _1$-norm and total variation (TV) of the brightness distribution.
In the proposed method, optimal regularization parameters, which represent the sparseness and effective spatial resolution of the image, are derived from data themselves using cross validation (CV).
As an application of this technique, we present simulated observations of M87 with the EHT based on four physically motivated models.
We confirm that $\ell _1$+TV regularization can achieve an optimal resolution of $\sim 20-30$\% of the diffraction limit $\lambda/D_{\rm max}$, which is the nominal spatial resolution of a radio interferometer.
With the proposed technique, the EHT can robustly and reasonably achieve super-resolution sufficient to clearly resolve the black hole shadow.
These results make it promising for the EHT to provide an unprecedented view of the event-horizon-scale structure in the vicinity of the super-massive black hole in M87 and also the Galactic center Sgr A*. 
\end{abstract}
\keywords{accretion, accretion disks --- black hole physics --- Galaxy: center --- submillimeter: general --- techniques: interferometric}

\section{Introduction}\label{sec:1}
Supermassive black holes (SMBHs) reside in the majority of the galactic nuclei in the universe. In a subset of such galaxies, accretion drives a highly energetic active galactic nucleus (AGN) often associated with powerful jets. Understanding the nature of these systems has been a central quest in astronomy and astrophysics. The SMBHs at the centers of our galaxy (Sgr A*) and the giant elliptical galaxy M87 provide unprecedented opportunities to directly image the innermost regions close to the central black hole, since the angular size of the event horizon is the largest among known black holes. The angular size of the Schwarzschild radius ($R_s$) is $\sim 10$~$\mu$as for Sgr A* for a distance of 8.3 kpc and a mass of $4.3\times 10^6$~$M_\odot$ \citep[e.g.][]{chatzopoulos2015}, and $\sim 3-7$ $\mu$as for M87 with a distance for 16.7~Mpc \citep{blakeslee2009} and a mass of $3-6 \times 10^{9}$~$M_{\odot}$ \citep[e.g.][]{gebhardt2011,walsh2013}. The apparent diameter of the dark silhouette of the black hole is $\sqrt{27}$~$R_s$ for the non-rotating black hole. It corresponds to $\sim 52$ $\mu$as for Sgr A* and $\sim 16-36$ $\mu$as for M87, which changes by only 4\% with the black-hole spin and viewing orientation \citep{bardeen1973}.

Very long baseline interferometric (VLBI) observations at short/sub-millimeter wavelengths ($\lambda \lesssim 1.3$~mm, $\nu \gtrsim 230$~GHz) can achieve a spatial resolution of a few tens of microarcseconds and therefore are expected to resolve event-horizon-scale structures, including the shadow of SMBHs. Indeed, recent significant progress on 1.3~mm VLBI observations with the Event Horizon Telescope \citep[EHT;][]{doeleman2009b} has succeeded in resolving compact structures of a few $R_s$ in the vicinity of the SMBHs in both Sgr A* and M87  \citep[e.g.][]{doeleman2008,doeleman2012,fish2011,fish2016,akiyama2015,johnson2015}. Direct imaging of these scales will be accessible in the next few years with technical developments and the addition of new (sub)millimeter telescopes such as the Atacama Large Submillimeter/millimeter Array (ALMA) to the EHT \citep[e.g.][]{fish2013}.

Regardless of the observing wavelength, angular resolution (often referred to as ``beam size'' in radio astronomy and ``diffraction limit'' in optical astronomy) is simply given by $\theta \approx \lambda/D_{\rm max}$, where $\lambda$ and $D_{\rm max}$ are the observed wavelength and the diameter of the telescope (or the longest baseline length for the radio interferometer), respectively. 
A practical limit for a ground-based, 1.3~mm VLBI array like the EHT is $\sim 25$ $\mu$as ($=1.3$~mm/$10000$~km), which is comparable to the radius of the black hole shadow in M87 and Sgr A*.
Hence, imaging techniques with good imaging fidelity at a spatial resolution smaller than $\lambda / D_{\rm max}$ would be desirable, particularly for future EHT observations of M87 and Sgr A*.

The imaging problem of interferometry is formulated as an under-determined linear problem when reconstructing the image from full-complex visibilities that are Fourier components of the source image. In the context of compressed sensing (also known as ``compressive sensing''), it has been revealed that an ill-posed linear problem may be solved accurately if the underling solution vector is sparse \citep{donoho2006,candes2006}. Since then, many imaging methods have been applied to radio interferometry \citep[see][and references therein]{garsden2015}. We call these approaches ``sparse modeling'' since they utilize the sparsity of the ground truth.

In \citet{honma2014}, we applied LASSO \citep[Least Absolute Shrinkage and Selection Operator;][]{tibshirani1996}, a technique of sparse modeling, to interferometric imaging. LASSO solves under-determined ill-posed problems by utilizing the $\ell _1$-norm (see \S\ref{sec:2.2} for details). Minimizing the $\ell _1$-norm of the solution reduces the number of non-zero parameters in the solution, equivalent to choosing a sparse solution. The philosophy of LASSO is therefore similar to that of the traditional CLEAN technique \citep{hogbom1974}, which favors sparsity in the reconstructed image and has been independently developed as Matching Pursuit \citep{mallat1993} in statistical mathematics for sparse reconstruction. In \citet{honma2014}, we found that LASSO can potentially reconstruct structure $\sim 4$ times finer than $\lambda/D_{\rm max}$. Indeed, it works well for imaging the black hole shadow for M87 with the EHT in simulations.


Our previous work \citep{honma2014} has three relevant issues. The first issue is reconstructing the image only from the visibility amplitudes and closure phases (see \S\ref{sec:2.1}), which have been the standard data products of EHT observations \citep{lu2012,lu2013,akiyama2015,wagner2015,fish2016} and optical/infrared interferometry. The algorithm of \citet{honma2014} is applicable only for full-complex visibilities, which are the usual data products from longer-wavelength radio interferometers. We have recently developed a fast and computationally cheap method to retrieve the visibility phases from closure phases \citep[PRECL;][]{ikeda2016}, which can reconstruct the black hole shadow of M87 combined with LASSO in simulated EHT observations. However, since the phase reconstruction in PRECL adopts a different prior on visibilities than LASSO, the resultant image may not be optimized well in terms of $\ell _1$-norm minimization and sparse image reconstruction. Another potential approach is to solve for the image and visibility phases with $\ell _1$-norm regularization simultaneously, enabling us to reconstruct the image with full advantage of the regularization function.

The second issue is that the $\ell _1$-norm regularization might not provide a unique solution and/or could reconstruct an image that is too sparse image if the number of pixels with non-zero brightness is not small enough compared to the number of pixels. This violates a critical assumption in techniques with $\ell _1$-norm regularization that the solution (i.e. the true image) should be sparse. Such a situation may occur for an extended source or also even for a compact source if the imaging pixel size is set to be much smaller than the size of the emission structure. Pioneering work has made use of transforms to wavelet or curvelet bases, in which the image can be represented sparsely \citep[e.g.][]{li2011,carrillo2012,carrillo2014,garsden2015,dabbech2015}.  As another strategy to resolve this potential issue, in \citet{honma2014}, we proposed to add another regularization, Total Variation \citep[TV; e.g.][]{rudin1992}, which is another popular regularization in sparse modeling. TV is a good indicator for sparsity of the image in its gradient domain instead of the image domain (see \S\ref{sec:2.2} for details) and it has been applied to astronomical imaging \citep[e.g.][]{wiaux2010,mcewen2011,carrillo2012,carrillo2014,uemura2015, chael2016} that includes optical interferometric imaging without the visibility phases (e.g. MiRA; \citealt{thiebaut2008}; and also see \citealt{thiebaut2013} for a review). TV regularization generally favors a smooth image (i.e. with larger effective resolution) but with a sharp edge, in contrast with maximum entropy methods \cite[MEM; e.g.][]{narayan1986}, which favor a smooth edge \citep[see e.g.][for comparison between TV and MEM]{uemura2015}. Inclusion of TV regularization enables reconstruction of an extended image while preserving sharp emission features preferred by $\ell _1$-norm regularization, thereby extending the class of objects where sparse modeling is applicable. Indeed, regularization with both the $\ell _1$-norm and TV has been shown to be effective for imaging polarization with full complex visibilities in our recent work  \citep{akiyama2017}.

An important detail is the determination of regularization parameters (e.g. weights on regularization functions), which is common in the vast majority of existing techniques. Since one can not know the true image of the source a priori, one should evaluate goodness-of-fitting and select appropriate regularization parameters from the data themselves. In well-posed problems, one can use statistical quantities considering residuals between data and models as well as model complexity to avoid over-fitting, such as reduced $\chi ^2$, the Akaike information criterion (AIC) and the Bayesian information criterion (BIC) using the degrees of freedom to constrain model complexity. However, for ill-posed problems like interferometric imaging, degrees of freedom can not be rigorously defined, preventing the use of such statistical quantities. 

In this paper, we propose a new technique to reconstruct images from interferometric data using sparse modeling. The proposed technique directly solves the image from visibility amplitudes and closure phases.
In addition to the $\ell _1$-norm (LASSO), we also utilize another new regularization term, TV, so that a high-fidelity image will be obtained even with a small pixel size and/or for extended sources.
Furthermore, we propose a method to determine optimal regularization parameters with cross validation (CV; see \S\ref{sec:2.3}), which can be applied to many existing imaging techniques.
As an example, in this paper, we applied our new technique to data obtained from simulated observations of M87 with the array of the EHT expected in Spring 2017. 

\section{The Proposed Method}\label{sec:2}
\subsection{A brief introduction of the closure phase}\label{sec:2.1}
A goal of radio and optical/infrared interferometry is to obtain the brightness distribution $I(x, y)$ of a target source at a wavelength $\lambda$ or a frequency $\nu$, where $(x, y)$ is a sky coordinate relative to a reference position so called the phase-tracking center. The observed quantity is a complex function called visibility $V(u, v)$, which is related to $I(x, y)$ by two-dimensional Fourier transform given by
\begin{equation}
V(u,v) = \int dxdy\,I(x,y) e^{-i2\pi(ux+vy)}.
\end{equation}
Here, the spatial frequency $(u,v)$ corresponds to the baseline vector (in units of the observing wavelength $\lambda$) between two antennas projected to the tangent plane of the celestial sphere at the phase-tracking center.

Observed visibilities are discrete quantities, and the sky image can be approximated by a pixellated version where the pixel size is much smaller than the nominal resolution of the interferometer. The image can therefore be represented as a discrete vector ${\bf I}$, related to the Stokes visibilities ${\bf V}$ by a discrete Fourier transform ${\bf F}$:
\begin{equation}
{\bf V}= {\bf F} {\bf I}. \label{eq:obseq}
\end{equation}
The sampling of visibilities is almost always incomplete. Since the number of visibility samples ${\bf V}$ is smaller than the number of pixels in the image, solving the above equation for the image ${\bf I}$ is an ill-posed problem. 

Here, we consider that the complex visibility $V_j$ is obtained from observation(s) with multiple antennas. Let us define its phase and amplitude as $\phi_j$ and $\bar{V}_j$, respectively, denoted as follows
\begin{eqnarray}
V_j = \bar{V}_j e^{i\phi_j},
\end{eqnarray}
where $j$ is the index of the measurement. Each measurement corresponds to a point $(u_j, v_j)$ in
$(u,v)$-plane and recorded at time $t_j$. In actual observations, some instrumental effects and the atmospheric turbulence primary from the troposphere induce the antenna-based errors in the visibility phase, leading that the observed phase $\tilde{\phi_j}$ is offset from the true phase $\phi_j$ of the true image. In particular, this is a serious problem in VLBI observations performed at different sites \citep[see][]{thompson2001}.

However, the robust interferometric phase information can be obtained through the measurements of the {\it closure phase}, defined as a combination of triple phases on a closed triangle of baselines recorded at the same time. It is known that the closure phase is free from antenna-based phase errors \citep{jennison1958}, which can be seen from the following definition of the closure phase,
\begin{equation}
\psi_m ^{123} = \tilde{\phi}_j ^{12} + \tilde{\phi}_k ^{23} + \tilde{\phi}_l ^{31} = \phi_j ^{12}+
\phi_k ^{23}+ \phi_l^{31},
\end{equation}
where $m$ is the index of the closure phase, and upper numbers ($1,2,3$) mean the index of stations involved in the closure phase or the visibility phase. The closure phase is also known as a phase term of the triple product of visibilities on closed baselines recorded at the same time, $V_j^{12}V_k^{23}V_l^{31}$, known as the {\it bi-spectrum}\footnote{Data products of visibility amplitudes and closure phases are also sometimes named as ``bi-spectrum'' in literature \citep[e.g.][]{buscher1994}. In this paper, we strictly distinguish them.}. Closure phases have been used to calibrate visibility phases in VLBI observations \citep[e.g.][]{rogers1974}.

In short/sub-millimeter VLBI or optical/infrared interferometry, the stochastic atmospheric turbulence in the troposphere over each station induces a rapid phase rotation in the visibility, making it difficult to calibrate or even measure the visibility phase reliably \citep[e.g., see][]{rogers1995,thiebaut2013}. Thus, image reconstruction using more robust closure phases, free from station-based phase errors, is useful for interferometric imaging with such interferometers.

\subsection{Image Reconstruction from Visibility Amplitudes and Closure Phases}\label{sec:2.2}
In this paper, we propose a method to solve the two-dimensional image ${\bf I}=\left\{I_{i,j}\right\}$ by the following equations:
\begin{equation}
\min _{\bf I} \,C({\bf I})\,\,\,{\rm subject\,\,\,to\,\,\,}I_{i,j}\geq0. \label{eq:equation}
\end{equation}
The cost function $C({\bf I})$ is defined as
\begin{equation}
C({\bf I})	=\chi^{2}({\bf I}) + \eta _{l} ||{\bf I}||_{1} + \eta _{t} ||{\bf I}||_{\rm tv}, \label{eq:cost_func}
\end{equation}
where $||{\bf I}||_{p}$ is $l_p$-norm of the vector ${\bf I}$ given by
\begin{equation}
||{\bf I}||_{p} = \left( \sum_i \sum_j |I_{i,j}|^p \right) ^{\frac{1}{p}}\,\,\,({\rm for}\,\,p>0),
\end{equation}
and $||{\bf I}||_{\rm tv}$ indicates an operator of TV.

The first term of Eq.(\ref{eq:cost_func}) is the traditional $\chi^{2}$ term representing the deviation between the reconstructed image and observational data (i.e. the visibility amplitude ${\bf \bar{V}}=\{|V_j|\}$ and closure phase ${\bf \Psi} = \{\psi_m\}$), defined by
\begin{equation}
\chi^{2}({\bf I}) = || {\bf \bar{V}} - {\bf A}({\bf F}{\bf I})||_2^2 + || {\bf \Psi} - {\bf B}({\bf F}{\bf I})||_2^2,   
\end{equation}
where ${\bf A}$ and ${\bf B}$ indicate operators to calculate the visibility amplitude and closure phase, respectively. Deviations between the model and observational data are normalized with the errors. This form of the residual sum of squares (RSS) is originally proposed in the Bi-spectrum Maximum Entropy Method \citep[BSMEM;][]{buscher1994} and also for modeling EHT data \citep[e.g.][]{lu2012,lu2013}. Note that it could be replaced to a RSS term for bi-spectra \citep[e.g.][]{bouman2015,chael2016}.

The second term represents LASSO-like regularization using the $\ell _1$-norm. Under the non-negative condition, $\ell _1$-norm is equivalent to the total flux. $\eta_{\rm l}$ is the regularization parameter for LASSO, adjusting the degree of sparsity by changing the weight of the $\ell _1$-norm penalty. In general, a large $\eta_{\rm l}$ prefers a solution with very few non-zero components, while $\eta_{\rm l}=0$ introduces no sparsity. In this paper, we use the normalized regularization parameter $\tilde{\eta}_{\rm l}$
defined by
\begin{equation}
\eta_{\rm l} \equiv \tilde{\eta}_{\rm l}(N_{\rm amp}+N_{\rm cphase})/\max ({\bf \bar{V}}),
\end{equation}
which is less affected by the number of visibility amplitude and closure phase data points, $N_{\rm amp}$ and $N_{\rm cphase}$, respectively, and also by the total flux density of the target source.

The third term is the TV regularization, defined by the sum of all differences of the brightness between adjacent image pixels. In this paper, we adopt a typical form for two-dimensional images \citep[][]{rudin1992} that has been used in astronomical imaging, \citep[e.g.][]{wiaux2010,mcewen2011,uemura2015,chael2016}, defined as
\begin{eqnarray}
||{\bf I}||_{\rm tv}=\sum _i \sum _j \sqrt{|I_{i+1,j} - I_{i,j}|^2 + |I_{i,j+1} - I_{i,j}|^2}.
\end{eqnarray}
TV is a good indicator of image sparsity in its gradient domain instead of the image domain. 
TV is highly affected by the effective spatial resolution of the image. 
For instance, an image with a small TV value has blocks of image pixels whose brightness are similar, since the image is sparse in the gradient domain.
The size of such blocks is equivalent to the effective spatial resolution, getting smaller for images with higher TV values. 
Thus, the regularization parameter $\eta_{\rm t}$ adjusts the effective spatial resolution of the reconstructed image.
In general, a larger (smaller) $\eta_{\rm t}$ prefers smoother (finer) distribution of power with less (higher) discreteness, leading to larger (smaller) angular resolution. 
In the present work, we use the normalized regularization parameter $\tilde{\eta}_{{\rm t}}$ defined by
\begin{equation}
\eta_{\rm t} \equiv \tilde{\eta}_{\rm t}(N_{\rm amp}+N_{\rm cphase})/4\max ({\bf \bar{V}}),
\end{equation}
similar to the LASSO term. A factor of 4 is based on a property of TV that takes a difference in the brightness to all four directions at each pixel.
Note that a major difference to maximum entropy methods, which also favor smooth images, is that TV regularization has a strong advantage in edge-preserving; strong TV regularization favors a piecewise smooth structure, but with clear and often sharp boundaries between non-emitting and emitting regions.

The problem described in Eq.(\ref{eq:equation}, \ref{eq:cost_func}) is non-linear minimum optimization.  In this work, we adopt a non-linear programming algorithm L-BFGS-B \citep[][]{byrd1995,zhu1997} that is an iterative method for solving bound-constrained nonlinear optimization problems. L-BFGS-B is one of the quasi-Newton methods that approximates the Broyden-Fletcher-Goldfarb-Shanno (BFGS) algorithm using a limited amount of computer memory. In L-BFGS-B, the cost function and its gradient are used to determine the next model parameters at each iterative process. We approximately set partial derivatives to 0 at non-differentiable points for both the $\ell _1$-norm and TV. 
The partial derivatives of $\chi^2$ are calculated numerically with central differences. 
We use the latest Fortran implementation of L-BFGS-B \citep[L-BFGS-B v3.0;][]{morales2011}. We note that the problem is non-convex as other imaging techniques using closure quantities \citep[e.g.][]{buscher1994,thiebaut2008,bouman2015,chael2016}, and a global solution is generally not guaranteed. 

\subsection{Determination of Imaging Parameters}\label{sec:2.3}
In the proposed method, the most important tuning parameters are the regularization parameters for the $\ell _1$-norm ($\tilde{\eta}_{\rm l}$) and TV ($\tilde{\eta}_{\rm t}$), which determine the sparseness and effective spatial resolution of the image, respectively. Smaller regularization parameters generally favor images with larger numbers of non-zero image pixels and more complex image structure, which could give better $\chi^{2}$ values by over-fitting. On the other hand, large regularization parameters provide images that are too simple and that do not fit the data well. To determine optimal parameters, we need to evaluate the goodness-of-fit using Occam's razor to prevent over-fitting. 

In this work, we adopt cross validation (CV) to evaluate goodness-of-fit. CV is a measure of the relative quality of the models for a given set of data. CV checks how the model will generalize to an independent data set by using separate datasets for fitting the model and for testing the fitted model. CV consists of three steps: (1) randomly partitioning a sample of data into complementary subsets, (2) performing the model fitting on one subset (called the {\it training set}), and (3) validating the analysis on the other subset (called the {\it validation set}).
To reduce variability, multiple rounds of cross-validation are performed using different partitions, and the validation results are averaged over the rounds. If the regularization parameters are too small, the established model from the training set would be over-fitted and too complicated, resulting in a large deviation in the validation set. On the other hand, if the regularization parameters are too large, the established model would be too simple and not well-fitted to the training set, also resulting in a large deviation in the validation set. Thus, reasonable parameters can be estimated by finding a parameter set that minimizes deviations (e.g. $\chi^{2}$) of the validation set.

In this work, we adopted 10-fold CV for evaluating the goodness-of-fit. The original data were randomly\footnote{In this work, subsamples are obtained with a uniform probability regardless of baselines following the most basic style of the CV. However, there could be more effective way of choosing subsamples for interferometric imaging. The optimum partitioning for CV is in the scope of our next studies in near future.} partitioned into 10 equal-sized subsamples. 9 subsamples were used in the image reconstruction as the training set, and the remaining single subsample was used as the validation set for testing the model using $\chi^{2}$. We repeated the procedure by changing the subsample for validation data 10 times, until all subsamples were used for both training and validation. The $\chi^{2}$ values of the validation data were averaged and then used to determine optimal tuning parameters.

An important advantage of this method compared with previously proposed methods is that it is applicable to any type of regularization functions and also imaging with multiple regularization functions. For instance, \citet{carrillo2012,carrillo2014} and subsequent work solve images by utilizing the $\ell _1$-norm on wavelet-transformed image or TV regularization alone. In this case, the parameter can be uniquely determined from the $\ell _2 $-norm of the estimated uncertainties on observational data \citep[see][for details]{carrillo2012}. However, it is not straightforward to extend the idea for the problems with multiple regularization functions. For another example, \citet{garsden2015} proposes another heuristic method to determine the regularization parameters on $\ell _1$-regularization on the wavelet/curvelet-transformed image by estimating its noise level on each scale, which is successful. However, the method would not work for all types of regularization functions.  On the other hand, CV is a general technique that can be applied to imaging with any other regularization functions or any combination of them in principle, which include MEM \citep[e.g.][]{buscher1994,chael2016} and patch priors \citep[e.g.][]{bouman2015}. This advantage is particularly important for Sgr A*, which needs to involve a regularization function to mitigate the interstellar scattering effects \citep[scattring optics;][]{johnson2016} in addition to the general regularization function(s) for imaging.

A relevant disadvantage of this method is its computational cost, since $n$-fold CV requires to reconstruct $(n+1)$ images for each set of regularization parameters. Recently, an accurate approximation of CV, which can be derived from a single imaging on full data set for each parameter set, has been proposed for imaging from full complex visibilities with $\ell _1$+TV regularizations \citep{obuchi2016a,obuchi2016b}. Future development of such heuristic approximations for other types of data and regularization functions could overcome this issue.

\section{Imaging Simulations}\label{sec:3}
\subsection{Physically Motivated Models}\label{sec:3.1}
In this paper, we adopt four physical models previously proposed for 1.3~mm emission on event-horizon scales.

The first model is a simple, but qualitatively correct, force-free jet model (hereafter BL09) in the magnetically dominated regime presented in \citet{broderick2009} and \citet{lu2014}. We adopted a model image presented in \cite{akiyama2015}, which is based on the model parameters fitted to the results of 1.3~mm observations with the EHT in \citet{doeleman2012} and the SED of M87 (Broderick et al. in preparation). The approaching jet is predominant for this model \citep[see][for more details]{akiyama2015}. 

The second and third models are based on results of GRMHD simulations presented in \citet{dexter2012}. We used the representative models DJ1 and J2, which are based on the same GRMHD simulation but with different energy and spatial distributions for radio-emitting leptons. The dominant emission region is the accretion flow in DJ1 and the counter jet in J2 illuminating the last photon orbit in J2. We adopt model images in \citet{akiyama2015}, where the position angle of the large-scale jet for models is adjusted to $-70^{\circ}$ inferred for M87 \citep[e.g.][]{hada2011}.

The last model is based on results of GRMHD simulations presented in
\citet{moscibrodzka2016}, which models M87 core emission as radiation
produced by the jet sheath. We use the image averaged for $\sim$ 3~months for our simulation (hereafter M16). The image has its dominant contribution from the counter jet illuminating the last photon orbit similar to J2 of \citet{dexter2012}, but the M16 model assumes energy distributions of leptons quite different from J2. We rotate the original model image of \cite{moscibrodzka2016} to adjust the position angle of the large-scale jet to $-70^{\circ}$.

\subsection{Simulated Observations}\label{sec:3.2}
We simulate observations with the EHT at 1.3~mm (230~GHz) using the MAPS (MIT Array Performance Simulator) package\footnote{\url{http://www.haystack.mit.edu/ast/arrays/maps/}} based on the above models. The simulated observations are performed for the array expected to comprise in Spring 2017.

We assume an array consisting of stations at 6 different sites: a phased array of the Submillimeter Array (SMA) antennas and the James Clerk Maxwell Telescope (JCMT) on Mauna Kea in Hawaii; the Arizona Radio Observatory's Submillimeter Telescope (ARO/SMT) on Mt.\ Graham in Arizona; the Large Millimeter Telescope (LMT) on Sierra Negra, Mexico; a phased array of the Atacama Large Millimeter/submillimeter Array (ALMA) in the Atacama desert, Chile; the Institut de Radioastronomie Millim\'{e}trique (IRAM) 30m telescope on Pico Veleta, Spain; and a single dish telescope of the Northern Extended Millimeter Array (NOEMA) in France. We adopt the system equivalent flux density (SEFD) of each station shown in Table \ref{tab:sefd} based on the proposer's guide of 1-mm VLBI observations in ALMA Cycle 4.

\begin{table}
	\centering
	\caption{Stations used in simulated observations}
	\label{tab:sefd}
	\begin{tabular}{lc}
		\hline \hline
		Telescope & SEFD (Jy)\\
		\hline
		Phased ALMA&100\\
		Phased SMA and JCMT & 4000\\
		LMT&1400\\
		IRAM 30m&1400\\
		NOEMA single dish&5200\\
		ARO/SMT&11000\\
		\hline
	\end{tabular}
\end{table}

The simulations assume a bandwidth of 3.5~GHz for Stokes $I$, which is half of the standard setting in ALMA Cycle 4\footnote{\url{https://science.nrao.edu/observing/call-for-proposals/1mm-vlbi-cycle4/}}. 
We assume a correlation efficiency of 0.7, including a quantization efficiency of 0.88 for 2-bit sampling and other potential losses such as bandpass effects and pointing errors.

\begin{figure}
	\centering
	\includegraphics[width=1.0\columnwidth]{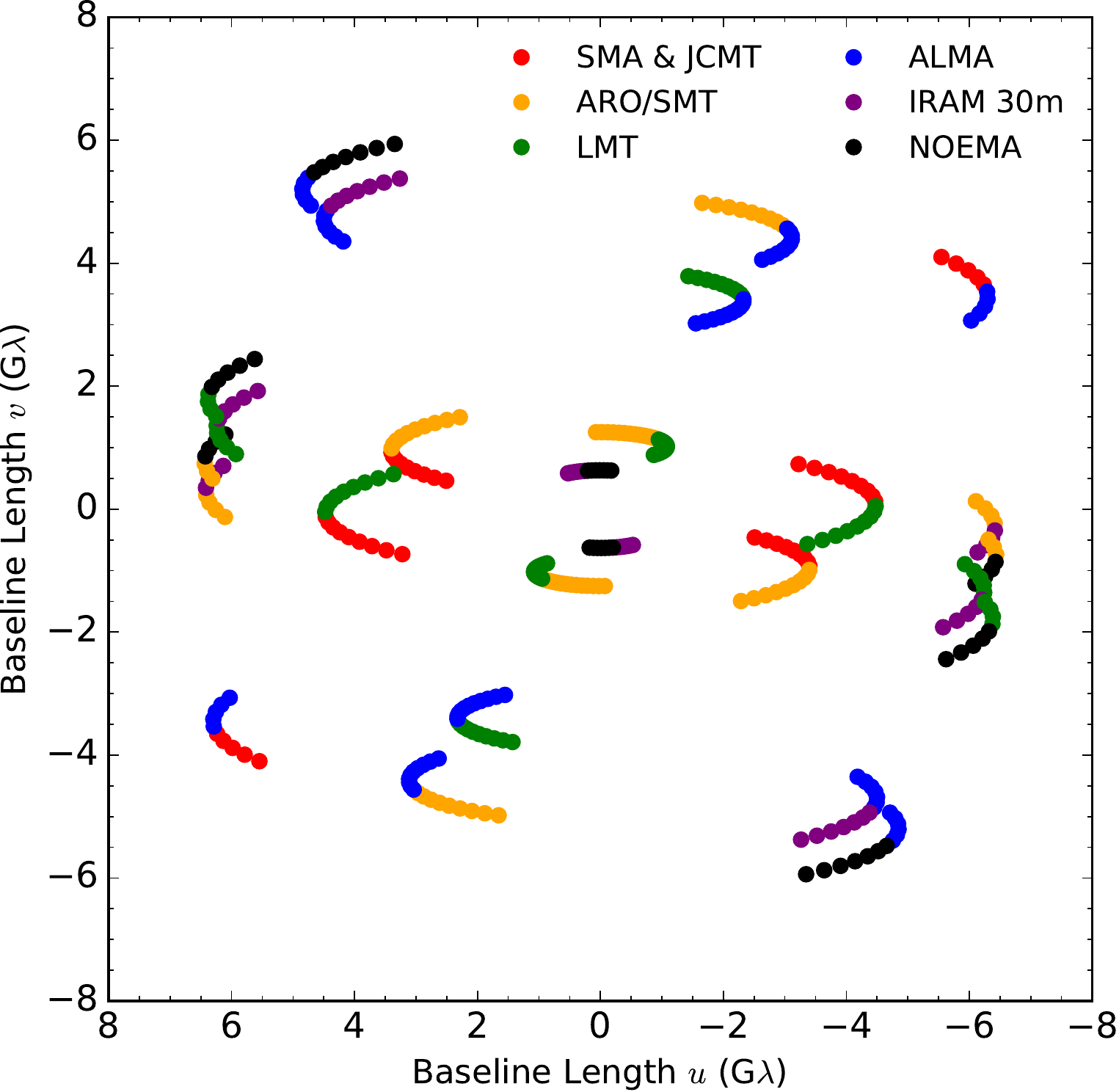}
	\caption{The $uv$-coverage of the simulated observations with the EHT array expected in Spring 2017. Each baseline is split in two colors to show involving stations.}
	\label{fig:uv-coverage}
\end{figure}

We simulate observations as a series of 5-minutes scans with a cadence of 20 minutes over a GST range of 13-0 hour, corresponding to the timerange when M87 can be observed by ALMA or LMT at an elevation greater than 20$^\circ$.  ALMA and LMT are sensitive stations near the middle of the east-west extent of the array, and they may be important anchor stations for fringe detection. This provides an observational efficiency of 25\% in time, expected for VLBI observations with ALMA in 2017. Data are integrated for the duration of each scan (i.e. 5 min) following previous EHT observations \citep[e.g.][]{doeleman2012,akiyama2015}. Figure~\ref{fig:uv-coverage} shows the resultant $uv$-coverage of simulated observations. The maximum baseline length of observations is 7.2~G$\lambda$, corresponding to $\lambda/D =28.5$~$\rm \mu$as.

We note that the conditions of our simulation are much worse than previous simulations in \citet{lu2014} in terms of the baseline sensitivity, $uv$-coverage, angular resolution (i.e. the maximum baseline length) and the exposure time of observations. Nevertheless, our simulated conditions are much closer to the observational conditions in Spring 2017.

\subsection{Imaging}\label{sec:3.3}
We reconstruct images from simulated data-sets based on the method described in \S\ref{sec:2.1}. We adopt a field of view (FOV) of 200~$\mu$as gridded by 100~pixels in both the RA and Dec directions for all models, giving a pixel size of $\sim$1.6~${\rm \mu}$as corresponding to a physical scale of $\sim 0.21$~$R_s$. Images are reconstructed at 4 regularization parameters for both $\tilde{\eta}_{\rm l}$ and $\tilde{\eta}_{t}$, ranging as $10^{-1},10^{0},...,10^{+2}$. As a result, we obtain $4\times4=16$~images for each model.

Since the problem described in \S\ref{sec:2.1} is non-convex, our algorithm may be trapped in a local minimum and therefore may end up at an initial-condition dependent solution, similar to other algorithms using techniques minimizing non-convex functions with gradient descent methods. To avoid this, we start reconstructing images at $\tilde{\eta}_{\rm l}=\tilde{\eta}_{\rm t}=10^{2}$, which is expected to derive the simplest image among parameters we adopt, assuming a point source as initial images. This is then used as the initial image at other values of the regularization parameters.

At each parameter, we do iterations until achieving convergence or 1000 iterations, and then filter the output image with a hard thresholding defined by
\begin{equation}
I_i^{\rm filtered} = 
\left\{
\begin{array}{ll}
0 & (|I_i|<t) \\
I_i & ({\rm otherwise})
\end{array}
\right.
,
\end{equation}
where $t$ is a threshold. We repeat this process until the normalized root mean square error (NRMSE) between the previous and latest filtered images becomes smaller than $1 \%$. NRMSE is defined by \citep[e.g.][]{chael2016}
\begin{equation}
	{\rm NRMSE}({\bf I},\,{\bf K}) = \sqrt{\frac{\sum_i |I_i-K_i|^2}{\sum_i |K_i|^2}},
\end{equation}
where ${\bf I}$ and ${\bf K}$ are the image to be evaluated and the reference image, respectively. We adopt the latest non-filtered image as the final product. Although this procedure makes the computational time longer, we found this works very well for avoiding convergence at some local minima. In this paper, we set 10\% of the peak flux as a threshold $t$. For the simulated observational data in this paper, it takes typically about several to ten minutes on a standard desktop computer with six intel core-i7 CPU cores to reconstruct an image at each set of two regularization parameters.

We evaluate the goodness-of-fit for each image and then selected the best-fit images with 10-fold CV as described in \S\ref{sec:2.3}. The quality of the reconstructed images is evaluated with the NRMSE. Since the all model images have finer resolutions with narrower FOVs than the reconstructed images, we calculated these metrics as follows. First, we adjusted the pixel size of the reconstructed image to that of the model image with bi-cubic spline interpolation. Second, we adjusted the position offsets between these two images so that the positions of their centers of mass coincide\footnote{Previous works \citep[e.g.][]{lu2014,fish2014} derived position offsets between the model and reconstructed images by taking cross correlations of the two images. However, we found that the position offsets derived from cross correlations induce an additional error in the metrics due to errors in position offsets. We found that the center of mass for the image is a better indicator of the position offsets than the cross-correlation minimum.}, because absolute positions cannot be defined from visibility amplitudes and closure phases alone. Finally, the metrics were evaluated. In addition to NRMSE, we also measure structural dissimilarity \citep[][]{wang2004} between the model and reconstructed images using the DSSIM metric adopted work by \citet{lu2014} and \citet{fish2014}. Since both metrics show similar trends, we show only the behavior of the NRMSE in the figures that follow.


\subsubsection{Imaging with the Cotton-Schwab CLEAN}
For evaluating the performance of our techniques, we also reconstructed images with the most widely-used Cotton-Schwab CLEAN \citep[henceforth CS-CLEAN;][]{schwab1984} implemented in the Common Astronomy Software Applications (CASA) package\footnote{\url{https://casa.nrao.edu/}} with uniform weighting. Since CLEAN requires complex visibilities, we adopted the simulated complex visibilities with thermal noises. We set a gain of 0.1 and a threshold of 0.08~mJy~beam$^{-1}$, comparable to the image sensitivity of simulated observations. Since the fast Fourier transform is often used in CLEAN, a very small FOV can require a grid size in $uv$-plane that is too large, which could cause additional deconvolution errors. Hence, we set 1024 pixels with the same pixel size in each axis for the entire map, and put a CLEAN box in the central 100$\times$100 pixels to put CLEAN components in the same region as other techniques. We use the model image instead of the CLEAN map for calculating metrics, since the residual map, which is generally added to the CLEAN map, cannot be calculated for the proposed method.




\section{Results}\label{sec:4}
\subsection{The best-fit images}\label{sec:4.1}
\begin{figure*}[t]
	\centering
	\includegraphics[width=0.8\textwidth]{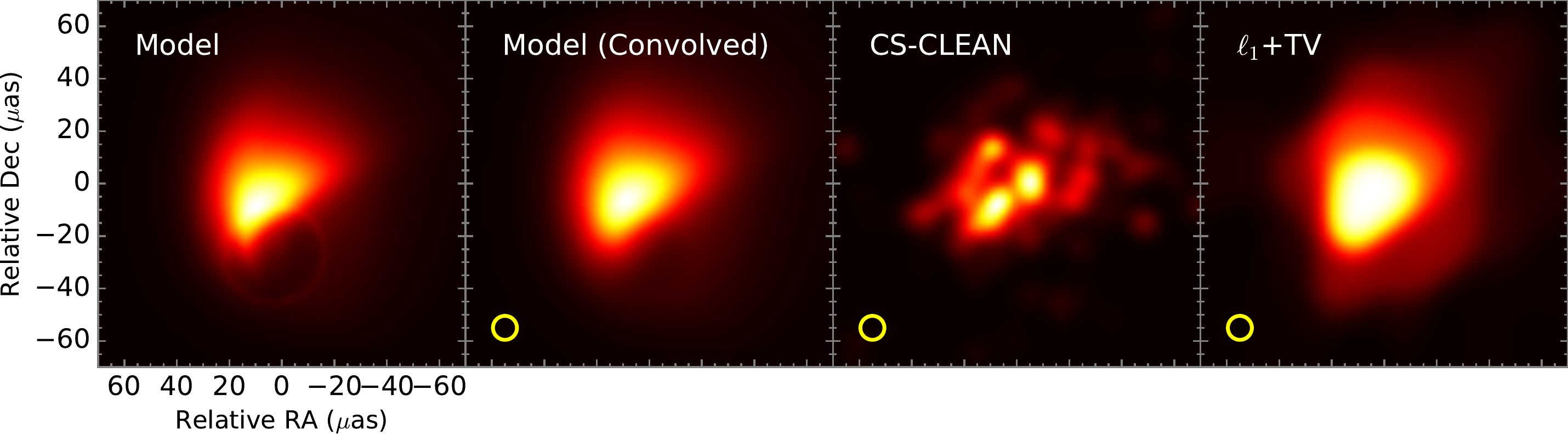}
	\includegraphics[width=0.8\textwidth]{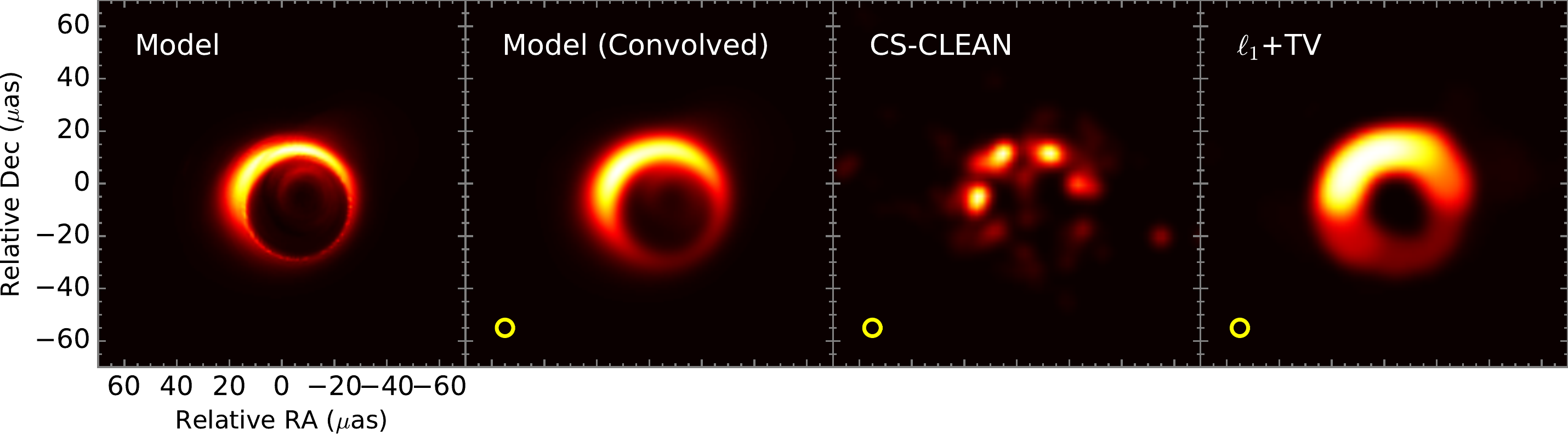}
	\includegraphics[width=0.8\textwidth]{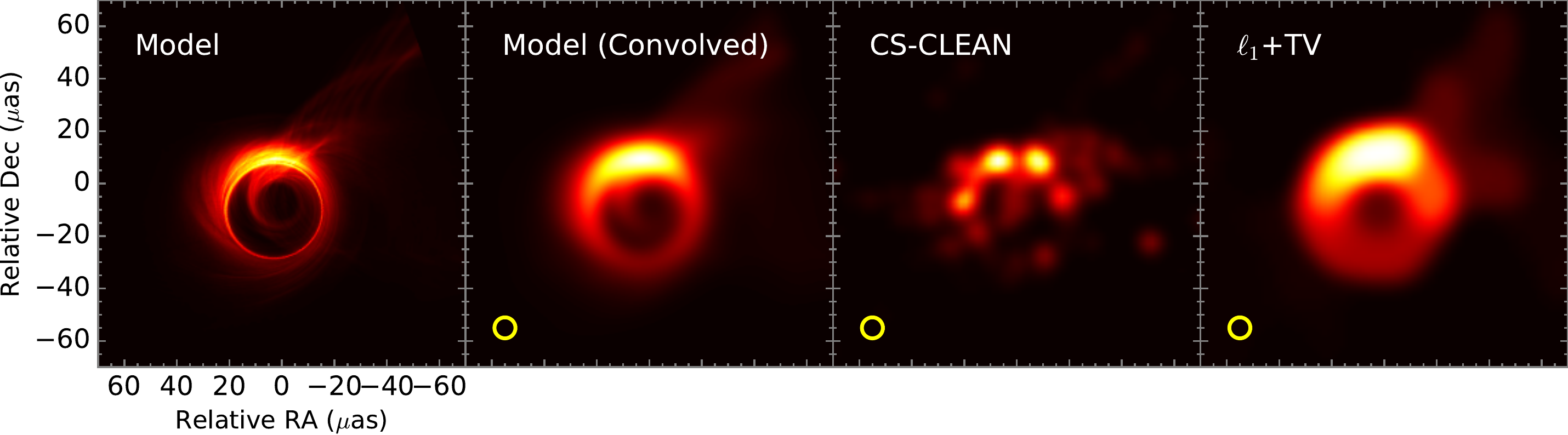}
	\includegraphics[width=0.8\textwidth]{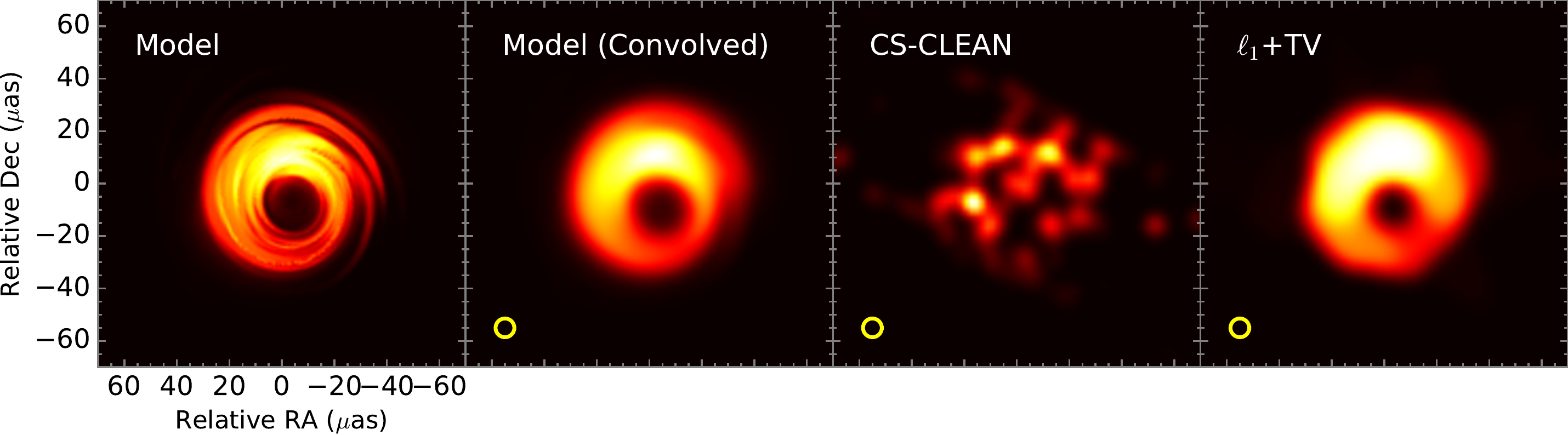}
	\caption{
		The model and reconstructed images. All images are convolved with circular Gaussian beams with the FWMH sizes corresponding to diameters of the yellow circles, which coincide with the optimal resolutions for $\ell _1$+TV regularization shown in Figure~\ref{fig:fidelity-resolution}.
		{\bf Top panels}: The approaching-jet-dominated model BL09 taken from \citet{akiyama2015} (originally from \citealt{broderick2009} and \citealt{lu2014}).
		{\bf The second and third panels from the top}: The counter-jet-dominated models J2 taken from \citet{akiyama2015} (originally from \citealt{dexter2012}) and M16 \citep{moscibrodzka2016}, respectively.
		{\bf Bottom panels}: The accretion-flow-dominated model DJ1 taken from \citet{akiyama2015} (originally from \citealt{dexter2012}).}%
	\label{fig:best-fit-images}
\end{figure*}

We show the best-fit images selected with CV in Figure~\ref{fig:best-fit-images}. A clear shadow feature is well reproduced for the counter-jet- and accretion-flow-dominated models (J2, M16 and DJ1).
This demonstrates that the EHT will achieve effective sufficient spatial resolution to image the black hole shadow of M87 if the mass-loading radius of the jet is not too large \citep{broderick2009,lu2014,akiyama2015}.

\begin{figure*}[t]
	\centering
	\includegraphics[width=1.0\textwidth]{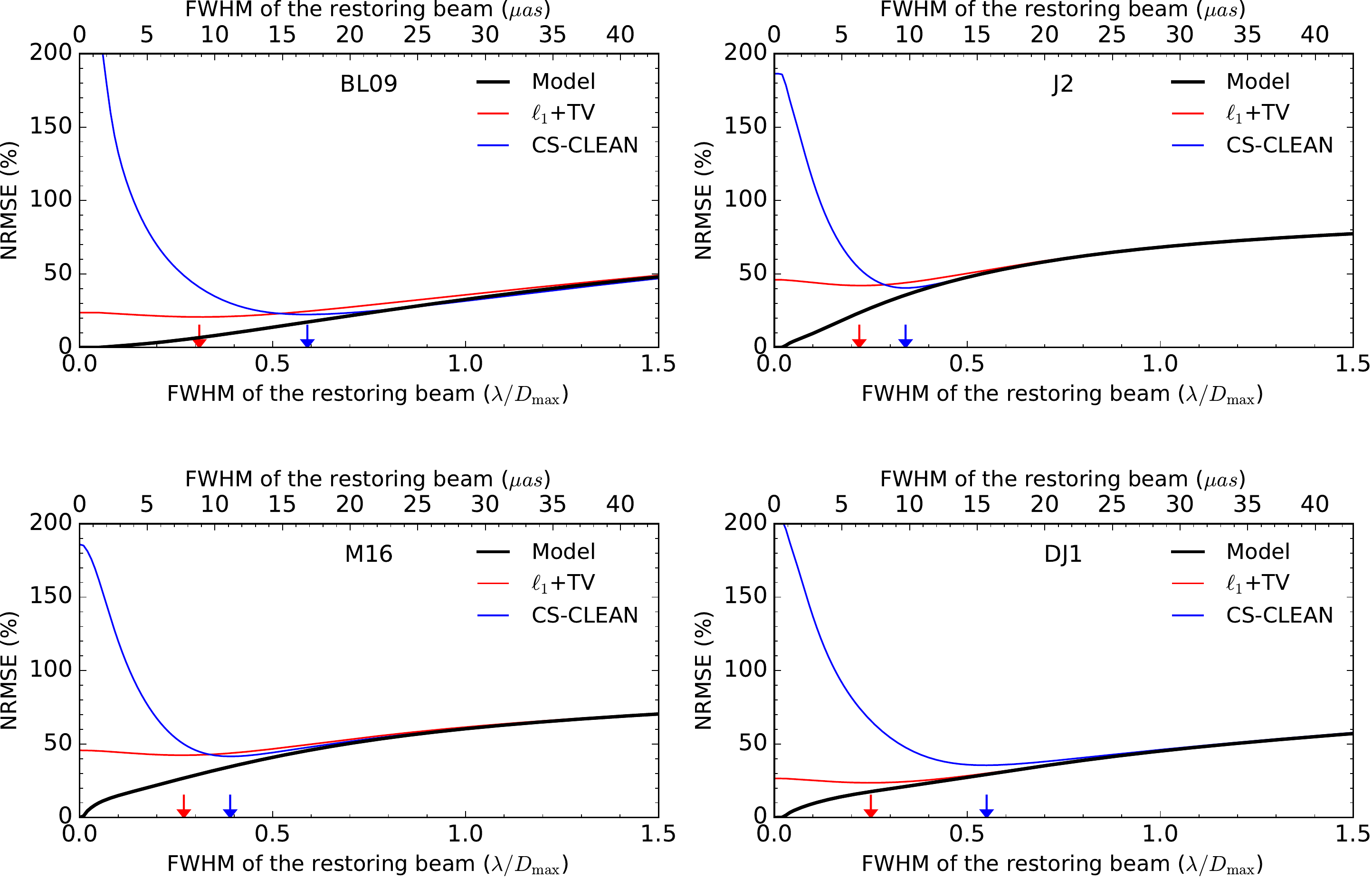}
	\caption{The NRMSE between the non-beam-convolved original model image and beam-convolved model/reconstructed images of all four models, as a function of the FWHM size of the convolving circular beam. The red and blue arrows indicate the optimal resolution of $\ell _1$+TV regularization and CS-CLEAN, respectively, which minimize the NRMSE.}
	\label{fig:fidelity-resolution}
\end{figure*}

Figure~\ref{fig:fidelity-resolution} shows the NRMSE metric for reconstructed images. The black curve labeled ``Model'' shows the NRMSE calculated when the model image is convolved with a circular Gaussian beam with a full width at half maximum (FWHM) as shown on the abscissa and compared against the original (unconvolved) model image. The Model curve effectively quantifies the best-case scenario in which the differences from the original input are due solely to a loss of resolution, not to errors in reconstructing the image. Figure~\ref{fig:fidelity-resolution} also show the NRMSE of each of the reconstructed images convolved with circular Gaussian beams.

Figure~\ref{fig:fidelity-resolution} clearly shows that closure-phase imaging with $\ell _1$+TV regularization works well compared to CS-CLEAN in particular at finer resolutions, despite the fact that CS-CLEAN uses full complex visibilities with more information and higher SNRs than closure phases. Both CS-CLEAN and $\ell _1$+TV images achieve similar NRMSEs on scales comparable to or greater than the diffraction limit. On the other hand, the NRMSEs of the reconstructed images start to deviate from the model images in the super-resolution regime---namely on scales smaller than the diffraction limit. In this regime, the NRMSEs differ by technique. CS-CLEAN has a common trend for all four models, which is broadly consistent with results of \citet{chael2016}. They achieve minimum errors at a resolution of $\sim 30-60$\% of the diffraction limit and then show a rapid increase in errors at smaller scales. In contrast, closure-phase imaging with $\ell _1$+TV regularizations show much more modest variations in the super-resolution regime. They achieve minimum errors at a resolution of $\sim 20-30$\% of the diffraction limit, smaller than CS-CLEAN, and show only a slight increase at smaller scales. $\ell _1$+TV reconstructions produce images that have a smooth distribution similar to the model images, resulting in smaller errors than CS-CLEAN, even if the $\ell _1$+TV reconstructions are not convolved with a restoring beam.

\begin{figure*}[t]
	\centering
	\includegraphics[width=0.7\textwidth]{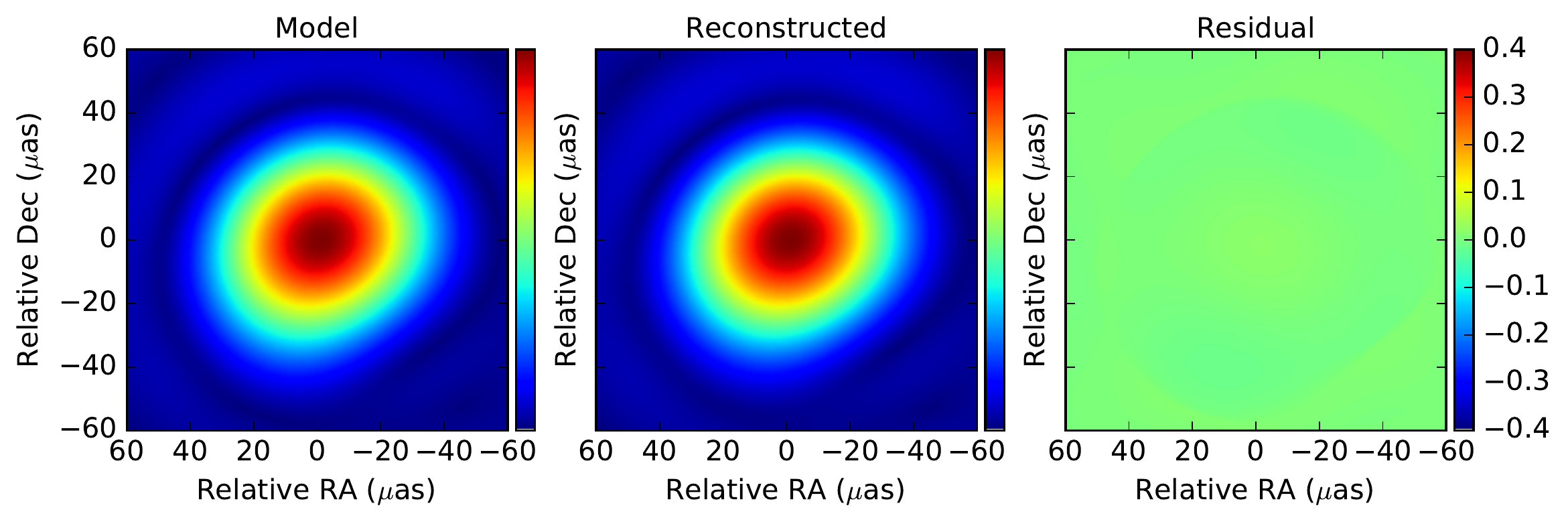}\\
	(a) Filtered with $|{\bf u}|_{\rm cutoff}=0.5D_{\rm max}/\lambda$\\
	\includegraphics[width=0.7\textwidth]{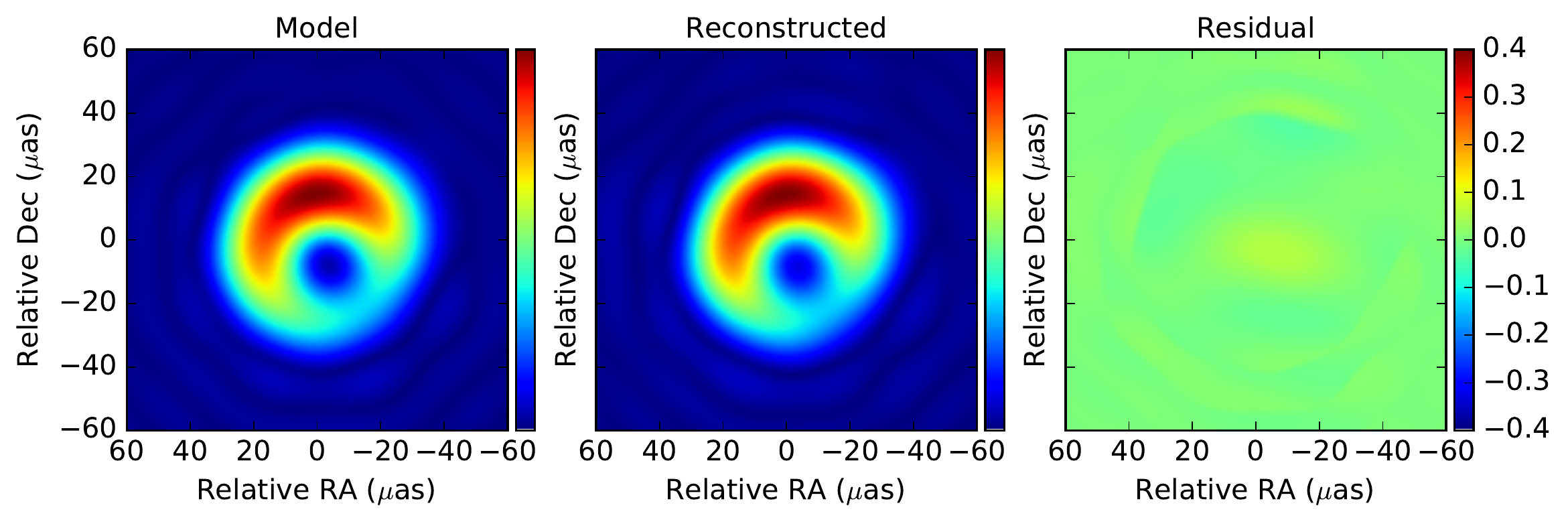}\\
	(b) Filtered with $|{\bf u}|_{\rm cutoff}=1.0D_{\rm max}/\lambda$\\
	\includegraphics[width=0.7\textwidth]{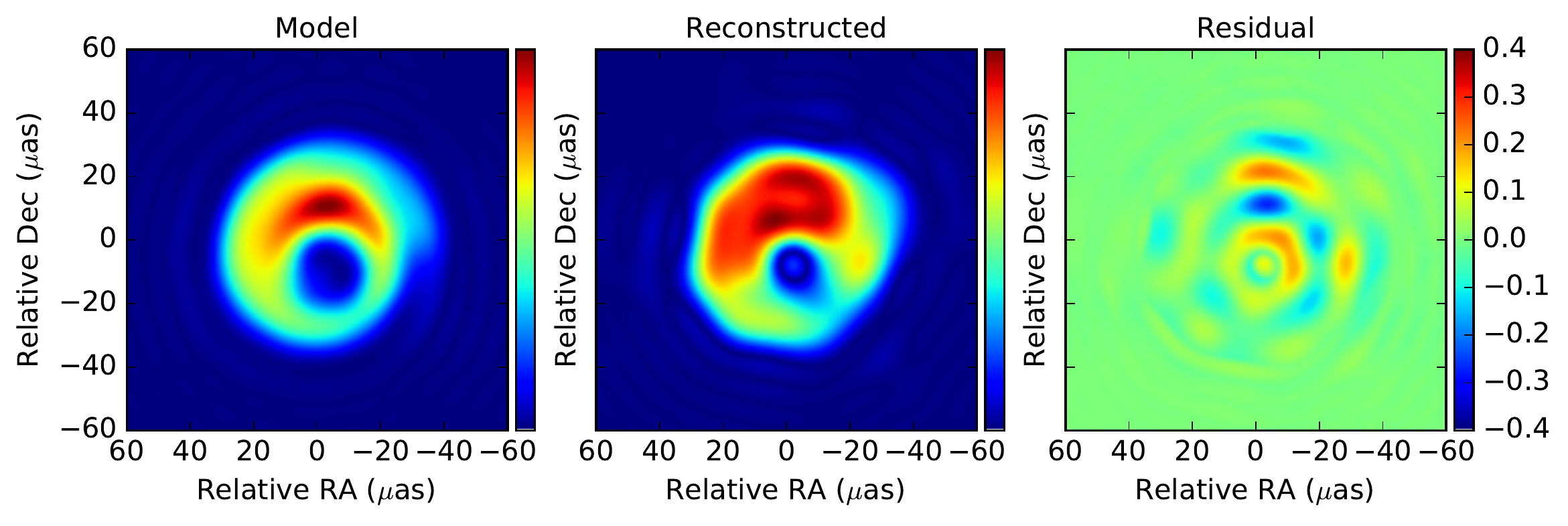}\\
	(c) Filtered with $|{\bf u}|_{\rm cutoff}=2.0D_{\rm max}/\lambda$\\
	\includegraphics[width=0.7\textwidth]{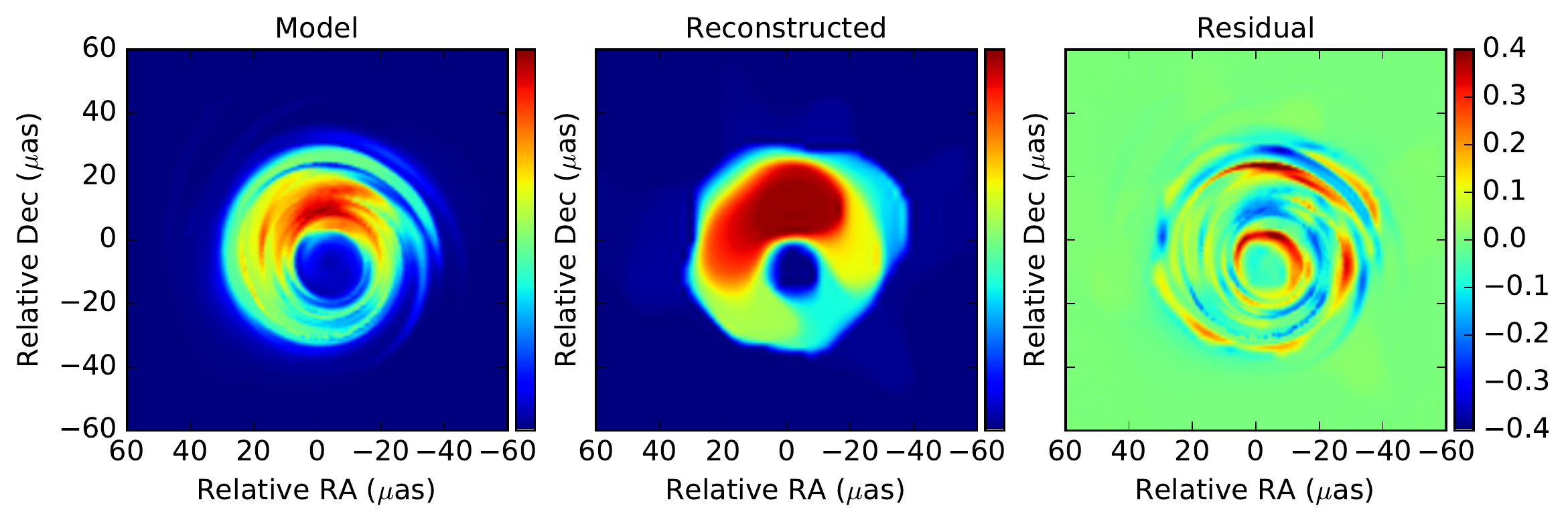}\\
	(d) Original Images (no filtering)\\
	\caption{The model, reconstructed, and residual images of DJ1 for different filtering baseline lengths. The left and middle panels show the model and best-fit images, while the right panels show the difference between these two images normalized with the peak brightness of the model image. Images at panels (a), (b) and (c) are low-pass filtered with spatial frequencies (or baseline lengths) of $|{\bf u}|_{\rm cutoff}=(0.5,\,1,\,2)\times D_{\rm max}/\lambda$, which are equivalent to convolving images with modified Bessel functions with FWHM sizes of $\sim (1.2,\,0.6,\,0.3)\times \lambda/D_{\rm max}$, respectively. The panels (d) show original images without filtering.}
	\label{fig:residuals}
\end{figure*}

In super-resolution regimes, the errors in $\ell _1$+TV images mostly arise from the presence of tiny substructures in the image. For instance, we show model, reconstructed and residual images filtered with different baseline lengths for DJ1 in Figure~\ref{fig:residuals}. As shown in Figure~\ref{fig:residuals}, residuals are small when filtering baseline lengths are shorter than the maximum baseline length. On the other hand, for longer filtering baseline lengths, systematic residuals due to tiny substructures much smaller than the diffraction limit start to appear, which can be traced only with baselines longer than the simulated observations. 

Although NRMSEs are better than CS-CLEAN at finer resolutions, $\ell _1$+TV images have broader emission region sizes regardless of models. This is due to a typical feature of images reconstructed with the isotropic TV, which prefers flat images with sharp edges. Since the simulated data do not have visibilities at baseline lengths long enough to resolve the width of ring- or crescent-like features, TV enlarges their widths until images start to deviate from observed visibilities. This property of the isotropic TV regularization would be useful to constrain the upper-limit size of the emission regions and black-hole shadow (see \S\ref{sec:5.1} for further discussions). On the other hand, in terms of the image fidelity, these results suggest that regularizations preferring much smoother edges are preferable; smoother gradients in the image lead to images with higher contrast (i.e. brighter/fainter pixels become even brighter/fainter, respectively) to conserve the total flux, which often makes the effective emission region size smaller. In \S\ref{sec:5.2}, we discuss alternative regularizations of sparse imaging reconstruction for improving the image fidelity in super-resolution regime.

\subsection{Regularization parameters and Cross Validation}\label{sec:4.2}
\begin{figure*}[t]
	\centering
	\includegraphics[width=0.7\textwidth]{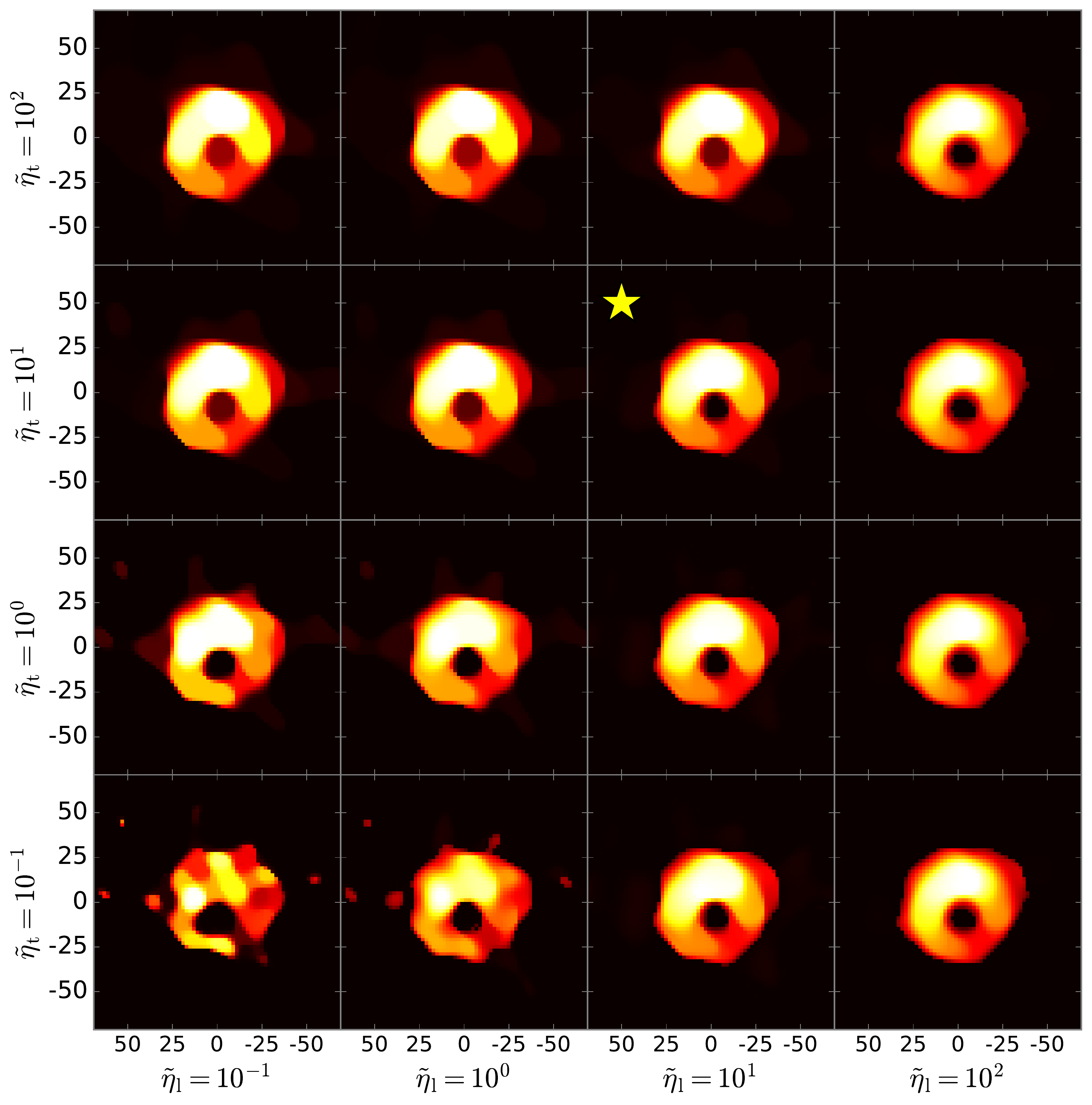}
	\caption{The parameter dependences of the reconstructed images in the regularization parameters for the accretion-flow-dominated model (DJ1). The yellow star indicates the regularization parameters for the best-fit images with the minimum CV. The units of the tick labels are $\mu$as.}%
	\label{fig:all-images}
\end{figure*}

\begin{figure*}[t]
	\centering
	\includegraphics[width=1.0\textwidth]{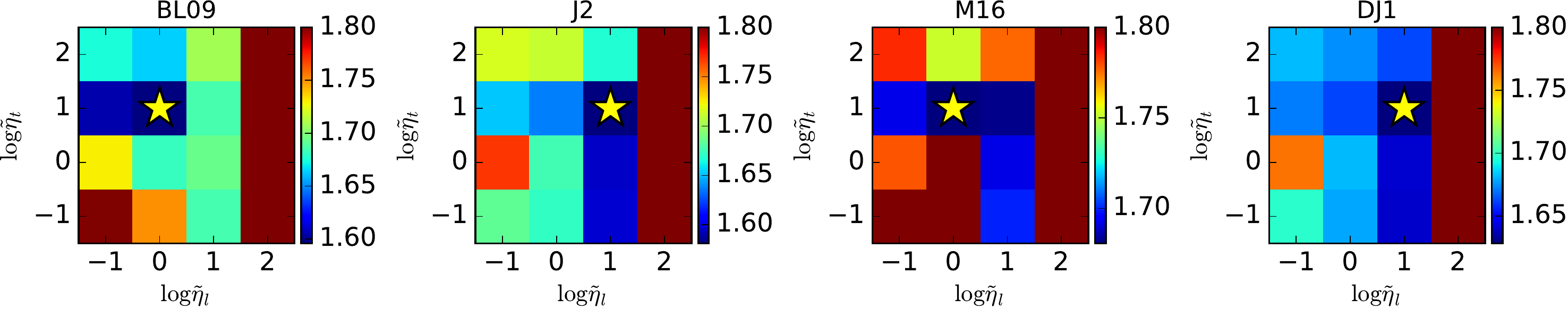}\\
	(a) Cross Validation ($\log {\rm CV}$)
	\includegraphics[width=1.0\textwidth]{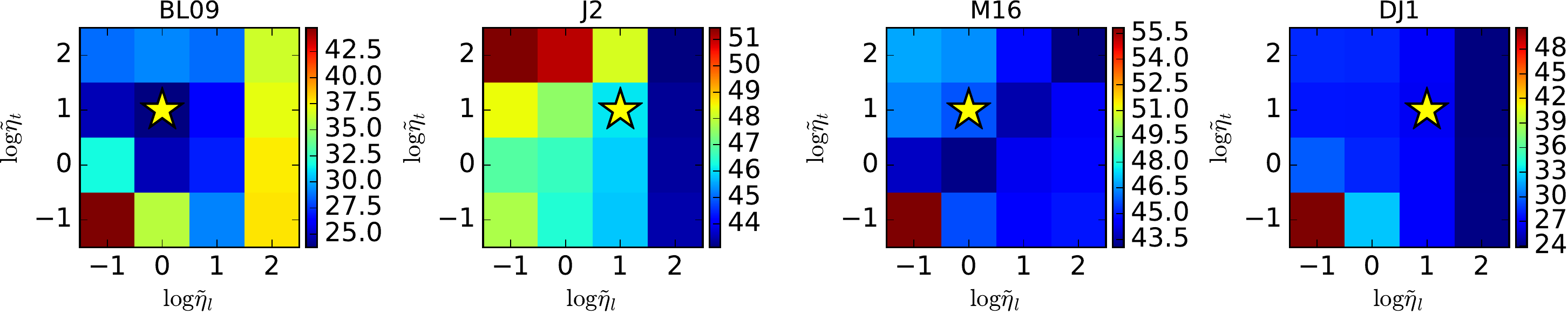}\\
	(b) NRMSE (\%)
	\caption{The logarithm of the CV value (a) and the NRMSE of non-Gaussian convolved image (b) for models BL09, J2, M16 and DJ1. The yellow stars indicate the regularization parameters for the best-fit images with the minimum CV value.  Note that,  in the panel (a), we set the upper limit of the color contours to values ($1.8$) lower than the maximum CV ($>2$ for all models) in order to highlight differences in CV around the best-fit parameters.}%
	\label{fig:param-dependence}
\end{figure*}

Reconstructed images for all 16 sets of the regularization parameters are shown in Figure~\ref{fig:all-images} for the accretion-flow-dominated model.
As shown in Figure~\ref{fig:all-images}, the reconstructed images with $\tilde{\eta}_{\rm l},\tilde{\eta}_{\rm t} \lesssim 1$ have noisy artifacts. This is true for all four models. Such artifacts may appear because the images are poorly constrained by both regularization functions at lower regularization parameters.

CV works as an Occam's razor and prevents over-fitting. In Figure~\ref{fig:param-dependence} (a), we show the residual $\chi ^2$ between the validation set and the model image reconstructed from the training set, which is averaged for all 10 trials (henceforth the CV value). 
As expected in \S\ref{sec:2.2}, the CV value tends to be large for large regularization parameters, since the regularization functions too strongly constrain the image so that the model image is inconsistent with observational data. 
Once the CV value achieves its minimum value (at a parameter set marked with stars in Figure~\ref{fig:param-dependence} (a)), it starts to increase again for lower regularization parameters, since the model image is over-fitted to the training set and then shows larger deviations from the validation set. 

Since the groundtruth images are known in this work, one can compare CV-selected images with images on other parameter sets, which can not be done for real observational data with unknown ground truth images. Although CV selects the optimal parameter based on the noise level of the data, it does not guarantee that the selected parameter achieves the best imaging fidelity among all parameter sets examined. In Fig \ref{fig:param-dependence} (b), we show the NRMSE of non-Gaussian-convolved images for all parameter sets. Although metrics for the image fidelity of best-fit images are slightly larger than the minimum values for the three of four models (J2, M16 and DJ1), they are consistent to within a few percents. These slight differences between the best-fit and best-fidelity parameters would not produce substantial differences in the image, and the resulting images are good enough.

The above results clearly demonstrate that CV is a useful technique to determine the regularization parameters so that the reconstructed image does not overfit noises in the data. We emphasize that CV is a general technique and can be applied to imaging regardless of the specific data products used (e.g. full complex visibilities, visibility amplitudes and/or closure quantities) or chosen regularization (for instance, sparse modeling; MEM:~e.g.,~\citealt{buscher1994},~\citealt{chael2016}; patch priors:~e.g.,~\citealt[][]{bouman2015}; scattering optics:~\citealt[][]{johnson2016}).

\section{Discussions}\label{sec:5}
\subsection{Implications for future EHT observations of M87}\label{sec:5.1}
The proposed method successfully reproduces a clear signature for the black hole shadow for the models J2, M16 and DJ1. This is as expected from the visibility distribution of these models, which have null amplitude regions, created by the shadow feature, at intermediate baseline lengths of $\sim 3-4$~G$\lambda$ \citep{akiyama2015}. MEM also succeeds in reproducing the black hole shadow in these models \citep{lu2014}. Presence of a clear shadow feature is tightly connected to where the dominant emission originates, since the silhouette of the black hole is created by the photons produced a few $R_s$ from the black hole, illuminating the last photon orbit \citep[see discussion in][]{akiyama2015}. As demonstrated in previous imaging simulations \citep{lu2014}, future EHT observations can constrain the loading radius of the high-energy leptons producing synchrotron emission at 1.3~mm via the appearance of the black hole.

The most important implication of this work is that our regularization function TV and parameter selection with CV will enlarge the ring- or crescent-like emission illuminating the last-photon orbit of the black hole as much as possible, within the range that the model image neither over-fits nor deviates too much from the observed visibilities. Hence, the obtained width of the surrounding emission in the reconstructed image is close to an upper-limit on the width of the emission region, and simultaneously, the obtained diameter of the black hole shadow should be interpreted as a reasonable lower limit for it. The clear shadow features in the reconstructed images for models J2, M16 and DJ1, therefore, strongly indicate that the EHT can sample a large enough range of visibilities with appropriately low noise levels to image the black hole shadow. In addition, the raw reconstructed image with the proposed method can be used to constrain the lower-limit size of the shadow. This would be useful to constrain the mass of M87, which has an uncertainty of an factor of about two between the stellar-dynamical \citep[e.g.][]{gebhardt2011} and gas-dynamical \citep[e.g.][]{walsh2013} modeling, and therefore be informative to clarify that which of modeling methods is desirable to measure the mass of super-massive black holes. 

Another important implication is that, at least for the black hole images, post-processing Gaussian convolution would not be required with the $\ell _1+TV$ regularization, although the CLEAN techniques do require it to reduce many compact artifacts in the image. As shown in Figure \ref{fig:fidelity-resolution}, the NRMSE curves for the $\ell _1+TV$ regularization are shallow for smaller convolving sizes, and applying a circular Gaussian beam therefore makes only small improvements of a few percent in the NRMSE regardless of the input model images. Similar results are also shown in recent work with the MEM \citep{chael2016}. Our results support that application of the beam is not required for the recent state-of-art imaging methods utilizing multi-resolution regularization functions for imaging the $R_s$-scale structure of M87 and Sgr A* with the EHT.

\subsection{Relevant future issues}\label{sec:5.2}
\subsubsection{Other sparse regularization for smoothed images}\label{sec:5.2.1}
In this paper, we adopt TV regularization, which favors a smooth image, so that images can be reconstructed with smaller pixel sizes and/or for more extended sources. As shown in $\S$\ref{sec:4}, the reconstructed images have good image fidelity. In particular, it is noteworthy that CV selected high values of $\tilde{\eta}_t = 10^{0}-10^{1}$ for all models (see Figure~\ref{fig:param-dependence} (a)), suggesting that the solutions would be over-fitted without TV regularization. These results demonstrate that inclusion of the TV regularization can extend the range of objects where sparse modeling is applicable.

However, as described in \S\ref{sec:4.1}, the reconstructed images have larger emission regions than the original model, since the isotropic TV \citep{rudin1992} adopted in this work prefers a flat brightness distribution in super-resolution regimes where the image cannot be constrained well. Hence, a regularization preferring more smoothed edges is required to improve the fidelity for the black-hole imaging with the EHT. 

In the context of sparse reconstruction, there are several candidates for improving the image fidelity as a natural extension of this work. First of all, there are other forms of regularization functions, which prefer sparse images in the gradient domain with smoother edges. For instance, an alternative form, given by,
\begin{eqnarray}
||{\bf I}||_{\rm tv^2}=\sum _i \sum _j \left(|I_{i+1,j} - I_{i,j}|^2 + |I_{i,j+1} - I_{i,j}|^2 \right).
\end{eqnarray}
is also convex like the isotropic TV term adopted in this work, and prefers images with smoother edges. Furthermore, previous studies of sparse image reconstruction techniques have shown that regularization with $\ell _1$+wavelet/curvelet transformation is also a promising approach \citep[e.g.][]{li2011,carrillo2012,carrillo2014,garsden2015,dabbech2015}. We will test these sparse regularizations in a forthcoming paper.

\subsubsection{Enhancing dynamic range with self-calibration}\label{sec:5.2.2}
In VLBI, the visibility phase is initially calibrated with fringe fitting (also called as fringe search), which is a self-calibration technique using phase closure \cite[see][]{thompson2001}. The fringe fitting can mitigate most station-based errors due to atmospheric and instrumental effects, although errors may remain if an incorrect source model is assumed. Traditionally, self-calibration with hybrid/differential mapping \citep[e.g.][]{walker1995} has been employed to solve for residual structural phase errors and images simultaneously, which has been successful for VLBI imaging.

This work and previous works on closure-phase imaging techniques using other regularizations such as MEMs \citep{lu2014,lu2016,fish2014,chael2016} and patch priors \citep[CHIRP;][]{bouman2015} demonstrate that an image can be reconstructed with high fidelity even from closure quantities. However, since closure phases and other closure quantities have less information about the source structure and also larger thermal noises than full complex visibilities, imaging with closure quantities can limit the dynamic range, image sensitivity and optimal spatial resolution.

A promising approach to improve the dynamic range is to use a reconstructed image from closure imaging techniques as an initial image for hybrid/differential mapping. It can also be used as a model for fringe fitting, as self-calibrated images are often applied to detect more fringes on faint sources \citep[e.g.][]{hada2016}. In a forthcoming paper, we will evaluate the performance of such a hybrid-mapping technique including closure-phase/full-closure imaging priors. We emphasize that, for this purpose, one does not need to reconstruct the image with pixels much smaller than scales where the brightness distribution cannot be constrained by data and therefore will not affect results of self-calibration.

\section{Conclusions}\label{sec:6}
We have presented a new imaging technique reconstructing images from visibility amplitudes and closure phases by utilizing two regularizations of sparse modeling: the $\ell _1$-norm and Total Variation (TV). Furthermore, we also propose a method to select optimal regularization parameters with cross validation (CV), which can be applied to most existing imaging algorithms. As an example, we applied our technique to simulated observations of M87 with the EHT at 1.3~mm. Here, we summarize our conclusions.
\begin{enumerate}
	\item We find that $\ell _1$+TV regularization can achieve an optimal resolution of $\sim 20-30$\% of the diffraction limit $\lambda/D_{\rm max}$, which is the nominal spatial resolution of a radio interferometer. This optimal resolution is better than that of the most-widely used Cotton-Schwab CLEAN, which uses full complex visibilities.
	\item We confirm that cross validation (CV) works as an Occam's razor and prevents over-fitting when selecting the optimal regularization parameters. CV is a general method that can be applied to interferometric imaging more generally, such as imaging with full-complex visibilities and/or using other regularizations.
	\item Using $\ell _1+$TV regularization, the reconstructed image maximizes the width of the emission region within the range that it neither over-fits nor deviates too strongly from the data. Hence, the clear reproduction of the black hole shadow in the reconstructed image suggests that future EHT observations will have the $uv$-coverage and sensitivity sufficient for imaging it. In addition, the reconstructed image will be able to constrain the sizes of the black hole shadow and surrounding emission region.
\end{enumerate}
Finally, we remark that all of above results demonstrate the clear promise of the EHT for providing an unprecedented view of the event-horizon-scale structure of the super-massive black hole in M87 and also the Galactic center Sgr A*.

\acknowledgements
We thank the anonymous referee for his/her useful and constructive suggestions to improve the paper.
K.A thanks members Dr.~Michael~D.~Johnson and Dr.~Lindy~Blackburn for many useful suggestions on this work. 
K.A. and this work are financially supported by the program of Postdoctoral Fellowships for Research Abroad at the Japan Society for the Promotion of Science. 
M.M. acknowledges support from the ERC Synergy Grant (Grant 610058). 
Event Horizon Telescope work at MIT Haystack Observatory and the Harvard Smithsonian Center for Astrophysics is supported by grants from the National Science Foundation (NSF; AST-1440254, AST-1614868) and through an award from the Gordon and Betty Moore Foundation (GMBF-3561). 
Work on sparse modeling and Event Horizon Telescope at the Mizusawa VLBI Observatory is financially supported by the MEXT/JSPS KAKENHI Grant Numbers 24540242, 25120007 and 25120008. 

\software{MAPS (\url{http://www.haystack.mit.edu/ast/arrays/maps/}), CASA}


\begin{thebibliography}{}
	\expandafter\ifx\csname natexlab\endcsname\relax\def\natexlab#1{#1}\fi
	
	\bibitem[{{Akiyama} {et~al.}(2015){Akiyama}, {Lu}, {Fish}, {Doeleman},
		{Broderick}, {Dexter}, {Hada}, {Kino}, {Nagai}, {Honma}, {Johnson}, {Algaba},
		{Asada}, {Brinkerink}, {Blundell}, {Bower}, {Cappallo}, {Crew}, {Dexter},
		{Dzib}, {Freund}, {Friberg}, {Gurwell}, {Ho}, {Inoue}, {Krichbaum},
		{Loinard}, {MacMahon}, {Marrone}, {Moran}, {Nakamura}, {Nagar}, {Ortiz-Leon},
		{Plambeck}, {Pradel}, {Primiani}, {Rogers}, {Roy}, {SooHoo}, {Tavares},
		{Tilanus}, {Titus}, {Wagner}, {Weintroub}, {Yamaguchi}, {Young}, {Zensus}, \&
		{Ziurys}}]{akiyama2015}
	{Akiyama}, K., {Lu}, R.-S., {Fish}, V.~L., {et~al.} 2015, \apj, 807, 150
	
	\bibitem[{{Akiyama} {et~al.}(2017){Akiyama}, {Ikeda}, {Pleau}, {Fish},
		{Tazaki}, {Kuramochi}, {Broderick}, {Dexter}, {Mo{\'s}cibrodzka},
		{Gowanlock}, {Honma}, \& {Doeleman}}]{akiyama2017}
	{Akiyama}, K., {Ikeda}, S., {Pleau}, M., {et~al.} 2017, \aj, in Press.,
	arXiv:1702.00424
	
	\bibitem[{{Bardeen}(1973)}]{bardeen1973}
	{Bardeen}, J.~M. 1973, in Black Holes (Les Astres Occlus), ed. C.~{Dewitt} \&
	B.~S. {Dewitt}, 215--239
	
	\bibitem[{{Blakeslee} {et~al.}(2009){Blakeslee}, {Jord{\'a}n}, {Mei},
		{C{\^o}t{\'e}}, {Ferrarese}, {Infante}, {Peng}, {Tonry}, \&
		{West}}]{blakeslee2009}
	{Blakeslee}, J.~P., {Jord{\'a}n}, A., {Mei}, S., {et~al.} 2009, \apj, 694, 556
	
	\bibitem[{{Bouman} {et~al.}(2015){Bouman}, {Johnson}, {Zoran}, {Fish},
		{Doeleman}, \& {Freeman}}]{bouman2015}
	{Bouman}, K.~L., {Johnson}, M.~D., {Zoran}, D., {et~al.} 2015, ArXiv e-prints,
	arXiv:1512.01413
	
	\bibitem[{{Broderick} \& {Loeb}(2009)}]{broderick2009}
	{Broderick}, A.~E., \& {Loeb}, A. 2009, \apj, 697, 1164
	
	\bibitem[{{Buscher}(1994)}]{buscher1994}
	{Buscher}, D.~F. 1994, in IAU Symposium, Vol. 158, Very High Angular Resolution
	Imaging, ed. J.~G. {Robertson} \& W.~J. {Tango}, 91
	
	\bibitem[{Byrd {et~al.}(1995)Byrd, Lu, Nocedal, \& Zhu}]{byrd1995}
	Byrd, R.~H., Lu, P., Nocedal, J., \& Zhu, C. 1995, SIAM Journal on Scientific
	Computing, 16, 1190
	
	\bibitem[{Candes \& Tao(2006)}]{candes2006}
	Candes, E.~J., \& Tao, T. 2006, Information Theory, IEEE Transactions on, 52,
	5406
	
	\bibitem[{{Carrillo} {et~al.}(2012){Carrillo}, {McEwen}, \&
		{Wiaux}}]{carrillo2012}
	{Carrillo}, R.~E., {McEwen}, J.~D., \& {Wiaux}, Y. 2012, \mnras, 426, 1223
	
	\bibitem[{{Carrillo} {et~al.}(2014){Carrillo}, {McEwen}, \&
		{Wiaux}}]{carrillo2014}
	---. 2014, \mnras, 439, 3591
	
	\bibitem[{{Chael} {et~al.}(2016){Chael}, {Johnson}, {Narayan}, {Doeleman},
		{Wardle}, \& {Bouman}}]{chael2016}
	{Chael}, A.~A., {Johnson}, M.~D., {Narayan}, R., {et~al.} 2016, ArXiv e-prints,
	arXiv:1605.06156
	
	\bibitem[{{Chatzopoulos} {et~al.}(2015){Chatzopoulos}, {Fritz}, {Gerhard},
		{Gillessen}, {Wegg}, {Genzel}, \& {Pfuhl}}]{chatzopoulos2015}
	{Chatzopoulos}, S., {Fritz}, T.~K., {Gerhard}, O., {et~al.} 2015, \mnras, 447,
	948
	
	\bibitem[{{Dabbech} {et~al.}(2015){Dabbech}, {Ferrari}, {Mary}, {Slezak},
		{Smirnov}, \& {Kenyon}}]{dabbech2015}
	{Dabbech}, A., {Ferrari}, C., {Mary}, D., {et~al.} 2015, \aap, 576, A7
	
	\bibitem[{{Dexter} {et~al.}(2012){Dexter}, {McKinney}, \& {Agol}}]{dexter2012}
	{Dexter}, J., {McKinney}, J.~C., \& {Agol}, E. 2012, \mnras, 421, 1517
	
	\bibitem[{{Doeleman} {et~al.}(2009){Doeleman}, {Agol}, {Backer}, {Baganoff},
		{Bower}, {Broderick}, {Fabian}, {Fish}, {Gammie}, {Ho}, {Honman},
		{Krichbaum}, {Loeb}, {Marrone}, {Reid}, {Rogers}, {Shapiro}, {Strittmatter},
		{Tilanus}, {Weintroub}, {Whitney}, {Wright}, \& {Ziurys}}]{doeleman2009b}
	{Doeleman}, S., {Agol}, E., {Backer}, D., {et~al.} 2009, in astro2010: The
	Astronomy and Astrophysics Decadal Survey
	
	\bibitem[{{Doeleman} {et~al.}(2008){Doeleman}, {Weintroub}, {Rogers},
		{Plambeck}, {Freund}, {Tilanus}, {Friberg}, {Ziurys}, {Moran}, {Corey},
		{Young}, {Smythe}, {Titus}, {Marrone}, {Cappallo}, {Bock}, {Bower},
		{Chamberlin}, {Davis}, {Krichbaum}, {Lamb}, {Maness}, {Niell}, {Roy},
		{Strittmatter}, {Werthimer}, {Whitney}, \& {Woody}}]{doeleman2008}
	{Doeleman}, S.~S., {Weintroub}, J., {Rogers}, A.~E.~E., {et~al.} 2008, \nat,
	455, 78
	
	\bibitem[{{Doeleman} {et~al.}(2012){Doeleman}, {Fish}, {Schenck}, {Beaudoin},
		{Blundell}, {Bower}, {Broderick}, {Chamberlin}, {Freund}, {Friberg},
		{Gurwell}, {Ho}, {Honma}, {Inoue}, {Krichbaum}, {Lamb}, {Loeb}, {Lonsdale},
		{Marrone}, {Moran}, {Oyama}, {Plambeck}, {Primiani}, {Rogers}, {Smythe},
		{SooHoo}, {Strittmatter}, {Tilanus}, {Titus}, {Weintroub}, {Wright}, {Young},
		\& {Ziurys}}]{doeleman2012}
	{Doeleman}, S.~S., {Fish}, V.~L., {Schenck}, D.~E., {et~al.} 2012, Science,
	338, 355
	
	\bibitem[{{Donoho}(2006)}]{donoho2006}
	{Donoho}, D.~L. 2006, IEEE Transactions on Information Theory, 52, 1289
	
	\bibitem[{{Fish} {et~al.}(2013){Fish}, {Alef}, {Anderson}, {Asada}, {Baudry},
		{Broderick}, {Carilli}, {Colomer}, {Conway}, {Dexter}, {Doeleman}, {Eatough},
		{Falcke}, {Frey}, {Gab{\'a}nyi}, {G{\'a}lvan-Madrid}, {Gammie}, {Giroletti},
		{Goddi}, {G{\'o}mez}, {Hada}, {Hecht}, {Honma}, {Humphreys}, {Impellizzeri},
		{Johannsen}, {Jorstad}, {Kino}, {K{\"o}rding}, {Kramer}, {Krichbaum},
		{Kudryavtseva}, {Laing}, {Lazio}, {Loeb}, {Lu}, {Maccarone}, {Marscher},
		{Mart'{\i}-Vidal}, {Martins}, {Matthews}, {Menten}, {Miller}, {Miller-Jones},
		{Mirabel}, {Muller}, {Nagai}, {Nagar}, {Nakamura}, {Paragi}, {Pradel},
		{Psaltis}, {Ransom}, {Rodr{\'{\i}}guez}, {Rottmann}, {Rushton}, {Shen},
		{Smith}, {Stappers}, {Takahashi}, {Tarchi}, {Tilanus}, {Verbiest},
		{Vlemmings}, {Walker}, {Wardle}, {Wiik}, {Zackrisson}, \&
		{Zensus}}]{fish2013}
	{Fish}, V., {Alef}, W., {Anderson}, J., {et~al.} 2013, ArXiv e-prints,
	arXiv:1309.3519
	
	\bibitem[{{Fish} {et~al.}(2011){Fish}, {Doeleman}, {Beaudoin}, {Blundell},
		{Bolin}, {Bower}, {Chamberlin}, {Freund}, {Friberg}, {Gurwell}, {Honma},
		{Inoue}, {Krichbaum}, {Lamb}, {Marrone}, {Moran}, {Oyama}, {Plambeck},
		{Primiani}, {Rogers}, {Smythe}, {SooHoo}, {Strittmatter}, {Tilanus}, {Titus},
		{Weintroub}, {Wright}, {Woody}, {Young}, \& {Ziurys}}]{fish2011}
	{Fish}, V.~L., {Doeleman}, S.~S., {Beaudoin}, C., {et~al.} 2011, \apjl, 727,
	L36
	
	\bibitem[{{Fish} {et~al.}(2014){Fish}, {Johnson}, {Lu}, {Doeleman}, {Bouman},
		{Zoran}, {Freeman}, {Psaltis}, {Narayan}, {Pankratius}, {Broderick}, {Gwinn},
		\& {Vertatschitsch}}]{fish2014}
	{Fish}, V.~L., {Johnson}, M.~D., {Lu}, R.-S., {et~al.} 2014, \apj, 795, 134
	
	\bibitem[{{Fish} {et~al.}(2016){Fish}, {Johnson}, {Doeleman}, {Broderick},
		{Psaltis}, {Lu}, {Akiyama}, {Alef}, {Algaba}, {Asada}, {Beaudoin},
		{Bertarini}, {Blackburn}, {Blundell}, {Bower}, {Brinkerink}, {Cappallo},
		{Chael}, {Chamberlin}, {Chan}, {Crew}, {Dexter}, {Dexter}, {Dzib}, {Falcke},
		{Freund}, {Friberg}, {Greer}, {Gurwell}, {Ho}, {Honma}, {Inoue}, {Johannsen},
		{Kim}, {Krichbaum}, {Lamb}, {Le{\'o}n-Tavares}, {Loeb}, {Loinard},
		{MacMahon}, {Marrone}, {Moran}, {Mo{\'s}cibrodzka}, {Ortiz-Le{\'o}n},
		{Oyama}, {{\"O}zel}, {Plambeck}, {Pradel}, {Primiani}, {Rogers}, {Rosenfeld},
		{Rottmann}, {Roy}, {Ruszczyk}, {Smythe}, {SooHoo}, {Spilker}, {Stone},
		{Strittmatter}, {Tilanus}, {Titus}, {Vertatschitsch}, {Wagner}, {Wardle},
		{Weintroub}, {Woody}, {Wright}, {Yamaguchi}, {Young}, {Young}, {Zensus}, \&
		{Ziurys}}]{fish2016}
	{Fish}, V.~L., {Johnson}, M.~D., {Doeleman}, S.~S., {et~al.} 2016, ArXiv
	e-prints, arXiv:1602.05527
	
	\bibitem[{{Garsden} {et~al.}(2015){Garsden}, {Girard}, {Starck}, {Corbel},
		{Tasse}, {Woiselle}, {McKean}, {van Amesfoort}, {Anderson}, {Avruch}, {Beck},
		{Bentum}, {Best}, {Breitling}, {Broderick}, {Br{\"u}ggen}, {Butcher},
		{Ciardi}, {de Gasperin}, {de Geus}, {de Vos}, {Duscha}, {Eisl{\"o}ffel},
		{Engels}, {Falcke}, {Fallows}, {Fender}, {Ferrari}, {Frieswijk}, {Garrett},
		{Grie{\ss}meier}, {Gunst}, {Hassall}, {Heald}, {Hoeft}, {H{\"o}randel}, {van
			der Horst}, {Juette}, {Karastergiou}, {Kondratiev}, {Kramer}, {Kuniyoshi},
		{Kuper}, {Mann}, {Markoff}, {McFadden}, {McKay-Bukowski}, {Mulcahy}, {Munk},
		{Norden}, {Orru}, {Paas}, {Pandey-Pommier}, {Pandey}, {Pietka}, {Pizzo},
		{Polatidis}, {Renting}, {R{\"o}ttgering}, {Rowlinson}, {Schwarz}, {Sluman},
		{Smirnov}, {Stappers}, {Steinmetz}, {Stewart}, {Swinbank}, {Tagger}, {Tang},
		{Tasse}, {Thoudam}, {Toribio}, {Vermeulen}, {Vocks}, {van Weeren},
		{Wijnholds}, {Wise}, {Wucknitz}, {Yatawatta}, {Zarka}, \&
		{Zensus}}]{garsden2015}
	{Garsden}, H., {Girard}, J.~N., {Starck}, J.~L., {et~al.} 2015, \aap, 575, A90
	
	\bibitem[{{Gebhardt} {et~al.}(2011){Gebhardt}, {Adams}, {Richstone}, {Lauer},
		{Faber}, {G{\"u}ltekin}, {Murphy}, \& {Tremaine}}]{gebhardt2011}
	{Gebhardt}, K., {Adams}, J., {Richstone}, D., {et~al.} 2011, \apj, 729, 119
	
	\bibitem[{{Hada} {et~al.}(2011){Hada}, {Doi}, {Kino}, {Nagai}, {Hagiwara}, \&
		{Kawaguchi}}]{hada2011}
	{Hada}, K., {Doi}, A., {Kino}, M., {et~al.} 2011, \nat, 477, 185
	
	\bibitem[{{Hada} {et~al.}(2016){Hada}, {Kino}, {Doi}, {Nagai}, {Honma},
		{Akiyama}, {Tazaki}, {Lico}, {Giroletti}, {Giovannini}, {Orienti}, \&
		{Hagiwara}}]{hada2016}
	{Hada}, K., {Kino}, M., {Doi}, A., {et~al.} 2016, \apj, 817, 131
	
	\bibitem[{{H{\"o}gbom}(1974)}]{hogbom1974}
	{H{\"o}gbom}, J.~A. 1974, \aaps, 15, 417
	
	\bibitem[{{Honma} {et~al.}(2014){Honma}, {Akiyama}, {Uemura}, \&
		{Ikeda}}]{honma2014}
	{Honma}, M., {Akiyama}, K., {Uemura}, M., \& {Ikeda}, S. 2014, \pasj, 66, 95
	
	\bibitem[{{Ikeda} {et~al.}(2016){Ikeda}, {Tazaki}, {Akiyama}, {Hada}, \&
		{Honma}}]{ikeda2016}
	{Ikeda}, S., {Tazaki}, F., {Akiyama}, K., {Hada}, K., \& {Honma}, M. 2016,
	\pasj, arXiv:1603.07078
	
	\bibitem[{{Jennison}(1958)}]{jennison1958}
	{Jennison}, R.~C. 1958, \mnras, 118, 276
	
	\bibitem[{{Johnson}(2016)}]{johnson2016}
	{Johnson}, M.~D. 2016, \apj, 833, 74
	
	\bibitem[{{Johnson} {et~al.}(2015){Johnson}, {Fish}, {Doeleman}, {Marrone},
		{Plambeck}, {Wardle}, {Akiyama}, {Asada}, {Beaudoin}, {Blackburn},
		{Blundell}, {Bower}, {Brinkerink}, {Broderick}, {Cappallo}, {Chael}, {Crew},
		{Dexter}, {Dexter}, {Freund}, {Friberg}, {Gold}, {Gurwell}, {Ho}, {Honma},
		{Inoue}, {Kosowsky}, {Krichbaum}, {Lamb}, {Loeb}, {Lu}, {MacMahon},
		{McKinney}, {Moran}, {Narayan}, {Primiani}, {Psaltis}, {Rogers}, {Rosenfeld},
		{SooHoo}, {Tilanus}, {Titus}, {Vertatschitsch}, {Weintroub}, {Wright},
		{Young}, {Zensus}, \& {Ziurys}}]{johnson2015}
	{Johnson}, M.~D., {Fish}, V.~L., {Doeleman}, S.~S., {et~al.} 2015, Science,
	350, 1242
	
	\bibitem[{{Li} {et~al.}(2011){Li}, {Cornwell}, \& {de Hoog}}]{li2011}
	{Li}, F., {Cornwell}, T.~J., \& {de Hoog}, F. 2011, \aap, 528, A31
	
	\bibitem[{{Lu} {et~al.}(2014){Lu}, {Broderick}, {Baron}, {Monnier}, {Fish},
		{Doeleman}, \& {Pankratius}}]{lu2014}
	{Lu}, R.-S., {Broderick}, A.~E., {Baron}, F., {et~al.} 2014, \apj, 788, 120
	
	\bibitem[{{Lu} {et~al.}(2012){Lu}, {Fish}, {Weintroub}, {Doeleman}, {Bower},
		{Freund}, {Friberg}, {Ho}, {Honma}, {Inoue}, {Krichbaum}, {Marrone}, {Moran},
		{Oyama}, {Plambeck}, {Primiani}, {Shen}, {Tilanus}, {Wright}, {Young},
		{Ziurys}, \& {Zensus}}]{lu2012}
	{Lu}, R.-S., {Fish}, V.~L., {Weintroub}, J., {et~al.} 2012, \apjl, 757, L14
	
	\bibitem[{{Lu} {et~al.}(2013){Lu}, {Fish}, {Akiyama}, {Doeleman}, {Algaba},
		{Bower}, {Brinkerink}, {Chamberlin}, {Crew}, {Cappallo}, {Dexter}, {Freund},
		{Friberg}, {Gurwell}, {Ho}, {Honma}, {Inoue}, {Jorstad}, {Krichbaum},
		{Loinard}, {MacMahon}, {Marrone}, {Marscher}, {Moran}, {Plambeck}, {Pradel},
		{Primiani}, {Tilanus}, {Titus}, {Weintroub}, {Wright}, {Young}, \&
		{Ziurys}}]{lu2013}
	{Lu}, R.-S., {Fish}, V.~L., {Akiyama}, K., {et~al.} 2013, \apj, 772,
	arXiv:1305.3359
	
	\bibitem[{{Lu} {et~al.}(2016){Lu}, {Roelofs}, {Fish}, {Shiokawa}, {Doeleman},
		{Gammie}, {Falcke}, {Krichbaum}, \& {Zensus}}]{lu2016}
	{Lu}, R.-S., {Roelofs}, F., {Fish}, V.~L., {et~al.} 2016, \apj, 817, 173
	
	\bibitem[{Mallat \& Zhang(1993)}]{mallat1993}
	Mallat, S.~G., \& Zhang, Z. 1993, IEEE Transactions on signal processing, 41,
	3397
	
	\bibitem[{{McEwen} \& {Wiaux}(2011)}]{mcewen2011}
	{McEwen}, J.~D., \& {Wiaux}, Y. 2011, \mnras, 413, 1318
	
	\bibitem[{Morales \& Nocedal(2011)}]{morales2011}
	Morales, J.~L., \& Nocedal, J. 2011, ACM Trans. Math. Softw., 38, 7:1
	
	\bibitem[{{Mo{\'s}cibrodzka} {et~al.}(2016){Mo{\'s}cibrodzka}, {Falcke}, \&
		{Shiokawa}}]{moscibrodzka2016}
	{Mo{\'s}cibrodzka}, M., {Falcke}, H., \& {Shiokawa}, H. 2016, \aap, 586, A38
	
	\bibitem[{{Narayan} \& {Nityananda}(1986)}]{narayan1986}
	{Narayan}, R., \& {Nityananda}, R. 1986, \araa, 24, 127
	
	\bibitem[{{Obuchi} {et~al.}(2016){Obuchi}, {Ikeda}, {Akiyama}, \&
		{Kabashima}}]{obuchi2016b}
	{Obuchi}, T., {Ikeda}, S., {Akiyama}, K., \& {Kabashima}, Y. 2016, ArXiv
	e-prints, arXiv:1611.07197
	
	\bibitem[{Obuchi \& Kabashima(2016)}]{obuchi2016a}
	Obuchi, T., \& Kabashima, Y. 2016, Journal of Statistical Mechanics: Theory and
	Experiment, 2016, 053304
	
	\bibitem[{{Rogers} {et~al.}(1995){Rogers}, {Doeleman}, \& {Moran}}]{rogers1995}
	{Rogers}, A.~E.~E., {Doeleman}, S.~S., \& {Moran}, J.~M. 1995, \aj, 109, 1391
	
	\bibitem[{{Rogers} {et~al.}(1974){Rogers}, {Hinteregger}, {Whitney},
		{Counselman}, {Shapiro}, {Wittels}, {Klemperer}, {Warnock}, {Clark}, \&
		{Hutton}}]{rogers1974}
	{Rogers}, A.~E.~E., {Hinteregger}, H.~F., {Whitney}, A.~R., {et~al.} 1974,
	\apj, 193, 293
	
	\bibitem[{Rudin {et~al.}(1992)Rudin, Osher, \& Fatemi}]{rudin1992}
	Rudin, L.~I., Osher, S., \& Fatemi, E. 1992, Physica D: Nonlinear Phenomena,
	60, 259
	
	\bibitem[{{Schwab}(1984)}]{schwab1984}
	{Schwab}, F.~R. 1984, \aj, 89, 1076
	
	\bibitem[{{Thi{\'e}baut}(2008)}]{thiebaut2008}
	{Thi{\'e}baut}, E. 2008, in \procspie, Vol. 7013, Optical and Infrared
	Interferometry, 70131I
	
	\bibitem[{{Thi{\'e}baut}(2013)}]{thiebaut2013}
	{Thi{\'e}baut}, {\'E}. 2013, in EAS Publications Series, Vol.~59, EAS
	Publications Series, ed. D.~{Mary}, C.~{Theys}, \& C.~{Aime}, 157--187
	
	\bibitem[{{Thompson} {et~al.}(2001){Thompson}, {Moran}, \&
		{Swenson}}]{thompson2001}
	{Thompson}, A.~R., {Moran}, J.~M., \& {Swenson}, Jr., G.~W. 2001,
	{Interferometry and Synthesis in Radio Astronomy, 2nd Edition}
	
	\bibitem[{{Tibshirani}(1996)}]{tibshirani1996}
	{Tibshirani}, R. 1996, Journal of the Royal Statistical Society. Series B
	(Methodological), 58, 267
	
	\bibitem[{{Uemura} {et~al.}(2015){Uemura}, {Kato}, {Nogami}, \&
		{Mennickent}}]{uemura2015}
	{Uemura}, M., {Kato}, T., {Nogami}, D., \& {Mennickent}, R. 2015, \pasj, 67, 22
	
	\bibitem[{{Wagner} {et~al.}(2015){Wagner}, {Roy}, {Krichbaum}, {Alef},
		{Bansod}, {Bertarini}, {G{\"u}sten}, {Graham}, {Hodgson}, {M{\"a}rtens},
		{Menten}, {Muders}, {Rottmann}, {Tuccari}, {Weiss}, {Wieching}, {Wunderlich},
		{Zensus}, {Araneda}, {Arriagada}, {Cantzler}, {Duran}, {Montenegro-Montes},
		{Olivares}, {Caro}, {Bergman}, {Conway}, {Haas}, {Johansson}, {Lindqvist},
		{Olofsson}, {Pantaleev}, {Buttaccio}, {Cappallo}, {Crew}, {Doeleman}, {Fish},
		{Lu}, {Ruszczyk}, {SooHoo}, {Titus}, {Freund}, {Marrone}, {Strittmatter},
		{Ziurys}, {Blundell}, {Primiani}, {Weintroub}, {Young}, {Bremer},
		{S{\'a}nchez}, {Marscher}, {Chilson}, {Asada}, \& {Inoue}}]{wagner2015}
	{Wagner}, J., {Roy}, A.~L., {Krichbaum}, T.~P., {et~al.} 2015, \aap, 581, A32
	
	\bibitem[{{Walker}(1995)}]{walker1995}
	{Walker}, R.~C. 1995, in Astronomical Society of the Pacific Conference Series,
	Vol.~82, Very Long Baseline Interferometry and the VLBA, ed. J.~A. {Zensus},
	P.~J. {Diamond}, \& P.~J. {Napier}, 247
	
	\bibitem[{{Walsh} {et~al.}(2013){Walsh}, {Barth}, {Ho}, \& {Sarzi}}]{walsh2013}
	{Walsh}, J.~L., {Barth}, A.~J., {Ho}, L.~C., \& {Sarzi}, M. 2013, \apj, 770, 86
	
	\bibitem[{Wang {et~al.}(2004)Wang, Bovik, Sheikh, \& Simoncelli}]{wang2004}
	Wang, Z., Bovik, A.~C., Sheikh, H.~R., \& Simoncelli, E.~P. 2004, IEEE TRANS.
	IMAGE PROCESSING, 13, 600
	
	\bibitem[{{Wiaux} {et~al.}(2010){Wiaux}, {Puy}, \& {Vandergheynst}}]{wiaux2010}
	{Wiaux}, Y., {Puy}, G., \& {Vandergheynst}, P. 2010, \mnras, 402, 2626
	
	\bibitem[{Zhu {et~al.}(1997)Zhu, Byrd, Lu, \& Nocedal}]{zhu1997}
	Zhu, C., Byrd, R.~H., Lu, P., \& Nocedal, J. 1997, ACM Trans. Math. Softw., 23,
	550
	
\end{thebibliography}

\end{document}